\documentclass{article}
\pdfoutput=1
\usepackage[utf8]{inputenc}
\usepackage[english]{babel}
\usepackage{graphics}
\usepackage{graphicx}
\usepackage{xcolor}
\usepackage{geometry}
\usepackage{chngcntr}
\usepackage{hyperref}
\newcommand{\BB}{\href{https://cds.cern.ch/record/2691414/files/1910.11775.pdf}{BB}}

\title{Report on the ECFA Early-Career Researchers Debate on the\\2020 European Strategy Update for Particle Physics}
\author{The ECFA Early-Career Researchers}
\date{February 6, 2020}

\begin{document}
\renewcommand{\abstractname}{Executive Summary}

\maketitle

\textbf{List of editors/organisers}\\
\small{
A. Bethani, Université catholique de Louvain, Chemin du Cyclotron 2, Louvain-la-Neuve, Belgium\\
E. Brondolin, CERN, Esplanade des Particules 1, Geneva, Switzerland\\
A. A. Elliot, Queen Mary University of London, Mile End Road, London, United Kingdom\\
J. Garc\'ia Pardi\~nas, Universit\"at Z\"urich, Winterthurerstrasse  190, Z\"urich, Switzerland\\
G. Gilles, Bergische Universit{\"a}t Wuppertal, Gaussstrasse 20, Wuppertal, Germany\\
L. Gouskos, CERN, Geneve 23, Geneva, Switzerland\\
E. Gouveia, LIP, Campus de Gualtar, Braga, Portugal\\
E. Graverini, \'{E}cole Polytechnique F\'{e}d\'{e}rale de Lausanne (EPFL), Cubotron, Lausanne, Switzerland\\
N. Hermansson-Truedsson, Lund University (Currently at Universit\"{a}t Bern), S\"{o}lvegatan 14A, Lund, Sweden\\
A. Irles, Universit{\'e} Paris-Saclay, CNRS/IN2P3, IJCLab, Orsay, France \\
H. Jansen, DESY, Notkestr. 85, Hamburg, Germany\\
K. H. Mankinen, Lund University, Professorsgatan 1, Lund, Sweden\\
E. Manoni, INFN Sezione di Perugia, Via A Pascoli, Perugia, Italy\\
A. Mathad, Universit\"at Z\"urich, Winterthurerstrasse  190, Z\"urich, Switzerland\\
J. McFayden, CERN, Esplanade des Particules 1, Geneva, Switzerland\\
M.  Queitsch-Maitland, CERN, Esplanade des Particules 1, Geneva, Switzerland\\
J. Rembser, CNRS/IN2P3, Institut Polytechnique de Paris, Ecole  Polytechnique, Palaiseau, France\\
E. T. J. Reynolds, University of Birmingham, Edgbaston, Birmingham, United Kingdom\\
R. Sch\"ofbeck, HEPHY, Nikolsdorfergasse 18, Vienna, Austria\\
P. Schwendimann, Paul Scherrer Institute, Forschungsstrasse 111, Villigen PSI, Switzerland\\
S. Sekmen, Kyungpook National University, 80 Daehak-ro Buk-gu, Daegu, Republic of Korea\\
P. Sznajder, National Centre for Nuclear Research (NCBJ), Pasteura 7, Warsaw, Poland\\
S. L. Williams, University of Cambridge, JJ Thomson Avenue, Cambridge, United Kingdom\\
D. Zanzi, CERN, Esplanade des Particules 1, Geneva, Switzerland\\

}

\normalsize

\textbf{List of authors}\\
\small{
N. Andari, CEA, Saclay, France\\
L. Apolinário, LIP, Avenida Prof. Gama Pinto 2, Lisbon, Portugal\\
K. Augsten, Czech Technical University in Prague, Brehova 7, Prague, Czech Republic\\
E. Bakos, Institute if Physics Belgrade,  Pregrevica 118, Belgrade, Serbia\\
I. Bellafont, CELLS-ALBA, Carrer de la Llum 2-26, Cerdanyola del Vallès, Spain\\
L. Beresford, The University of Oxford, Keble road, Oxford, United Kingdom\\
A. Bethani, Université catholique de Louvain, Chemin du Cyclotron 2, Louvain-la-Neuve, Belgium\\
J. Beyer, DESY Hamburg, Notkestr. 85, Hamburg, Germany\\
L. Bianchini, INFN Sezione di Pisa, Edificio C - Polo Fibonacci Largo B. Pontecorvo, 3, Pisa, Italy\\
C. Bierlich, Lund University, S\"olvegatan 14A, Lund, Sweden\\
B. Bilin, ULB/IIHE, Boulevard du Triomphe, Brussels, Belgium\\
K. L. Bj{\o}rke,  University of Oslo, Sem S{\ae}lands vei 24, Oslo, Norway\\
E. Bols,  Vrije Universiteit Brussel, Pleinlaan 2, Brussels, Belgium\\
P. A. Brás, Laboratory of Instrumentation and Experimental Particle Physics, Av. Gama Pinto, Lisbon, Portugal\\
L. Brenner, DESY - Deutsches Elektronen-SYnchrotron, Notkestra\ss e 85, Hamburg, Germany\\
E. Brondolin, CERN, Esplanade des Particules 1, Geneva, Switzerland\\
P. Calvo, Centro de Investigaciones Energéticas, Medioambientales y Tecnológicas (CIEMAT), Avenida Complutense 40, Madrid, Spain\\
B. Capdevila, Institut de Física d'Altes Energies, Campus UAB, Edifici Cn, Bellaterra (Barcelona), Spain\\
I. Cioara, Horia Hulubei National Institute for R\&D in Physics and Nuclear Engineering, Reactorului 30 , Bucharest-MG, Romania \\
L. N. Cojocariu, Horia Hulubei National Institute for R\&D in Physics and Nuclear Engineering (IFIN-HH), Reactorului 30, Bucharest-MG, Romania\\
F. Collamati, Istituto Nazionale di Fisica Nucleare, Sezione di Roma, Piazzale Aldo Moro 2, Roma, Italy\\
A. de Wit, DESY, Notkestr. 85, Hamburg, Germany\\
F. Dordei, Istituto  Nazionale  di  Fisica  Nucleare  (INFN), sezione di Cagliari, Complesso  Universitario  di  Monserrato  -  S.P.  per  Sestu  Km  0.700, Monserrato (Cagliari), Italy\\
M. Dordevic, VINCA Institute of Nuclear Sciences, University of Belgrade, Mike Petrovica Alasa 12-14, Belgrade, Serbia\\
T. A. du Pree, Nikhef, National Institute of Subatomic Physics, Science Park 105, Amsterdam, The Netherlands\\
L. Dufour, CERN, Esplanade des Particules 1, Geneva, Switzerland\\
A. Dziurda, Institute of Nuclear Physics Polish Academy of Science, Radzikowskiego 152, Krakow, Poland\\
U. Einhaus, DESY, Notkestr. 85, Hamburg, Germany\\
A. A. Elliot, Queen Mary University of London, Mile End Road, London, United Kingdom\\
S. Esen, Nikhef, Science Park 105, Amsterdam, Netherlands \\
J. Ferradas Troitino, CERN, Esplanade des Particules 1, Geneva, Switzerland\\
C. Franco, LIP, Av. Prof. Gama Pinto 2, Lisbon, Portugal\\
J. Garc\'ia Pardi\~nas, Universit\"at Z\"urich, Winterthurerstrasse  190, Z\"urich, Switzerland\\
A. Garc\'ia Alonso, IFCA, Av. de los Castros, Santander, Spain\\
A. Ghosh, Universit{\'e} Paris-Saclay, CNRS/IN2P3, IJCLab, Orsay, France \\
G. Gilles, Bergische Universit{\"a}t Wuppertal, Gaussstrasse 20, Wuppertal, Germany\\
A. Giribono, INFN - LNF, Via Enrico Fermi 40, Frascati, Italy\\
L. Gouskos, CERN, Geneve 23, Geneva, Switzerland\\
E. Gouveia, LIP, Campus de Gualtar, Braga, Portugal\\
E. Graverini, \'{E}cole Polytechnique F\'{e}d\'{e}rale de Lausanne (EPFL), Cubotron, Lausanne, Switzerland\\
J. K. Heikkil{\"a}, University of Zurich, Winterthurerstrasse 190, Zurich, Switzerland\\
H. N. Heracleous, CERN, Esplanade des Particules 1, Geneva, Switzerland\\
T. Herman, Czech Technical University in Prague, Brehova 7, Prague, Czech Republic\\
N. Hermansson-Truedsson, Lund University (Currently at Universit\"{a}t Bern), S\"{o}lvegatan 14A, Lund, Sweden\\
J. Hrt\'{a}nkov\'{a}, Nuclear Physics Institute of the Czech Academy of Sciences, Husinec - \v{R}e\v{z} 130, \v{R}e\v{z}, Czech Republic\\
P. S. Hussain, HEPHY, Apostelgasse 23, Wien, Austria\\
A. Irles, Universit{\'e} Paris-Saclay, CNRS/IN2P3, IJCLab, Orsay, France \\
H. Jansen, DESY, Notkestr. 85, Hamburg, Germany\\
P. Kalaczynski, NCBJ, Pasteura 7, Warsaw, Poland\\
J. Karancsi, Institute for Nuclear Research (ATOMKI), Bem tér 18/c, Debrecen, Hungary\\
P. Kontaxakis, National and Kapodistrian University of Athens, Panepistimioupolis, Athens, Greece\\
S. Kostoglou, National Technical University of Athens (NTUA), Iroon Polytechniou 9, Athens, Greece\\
A. Koulouris, NTU Athens, 9, Iroon Polytechniou str, Athens, Greece\\
M. Koval, Charles University, V Holesovickach 2, Prague, Czech Republic\\
K. Krizkova Gajdosova, Czech Technical University in Prague, Brehova 7, Prague, Czech Republic\\
J. A. Krzysiak, IFJ PAN (Institute of Nuclear Physics, Polish Academy of Sciences), ul. Radzikowskiego 152, Krakow, Poland\\
M. Kuich, University of Warsaw, Pateura 5, Warsaw, Poland\\
O. Lantwin, Universit\"at Z\"urich, Winterthurerstrasse 190, Z\"urich, Switzerland\\
F. Lasagni Manghi, INFN Bologna,  Viale Carlo Berti Pichat 6, Bologna, Italy\\
L. Lechner, Institute for High Energy Physics, Nikolsdorfer Gasse 18, Vienna, Austria\\
S. Leontsinis, University of Zurich, Winterthurerstrasse 190, Zurich, Switzerland\\
K. Lieret, Ludwig Maximilan University, Geschwister-Scholl-Platz 1, Munich, Germany\\
A. Lobanov, Ecole Polytechnique, 91128 Palaiseau Cedex, Palaiseau, France\\
J. M. Lorenz, LMU Muenchen, Am Coulombwall 1, Garching, Germany\\
G. Luparello, Istituto Nazionale di Fisica Nucleare (INFN), sezione di Trieste, Via Alfonso Valerio 1, Trieste, Italy\\
N. Lurkin, University of Birmingham, University of Birmingham, Birmingham, United Kingdom\\
K. H. Mankinen, Lund University, Professorsgatan 1, Lund, Sweden\\
E. Manoni, INFN Sezione di Perugia, Via A Pascoli, Perugia, Italy\\
L. Mantani, Universit\'e catholique de Louvain, Chemin du Cyclotron 2, Louvain la Neuve, Belgium\\
R. Marchevski, CERN, Esplanade des Particules 1, Geneva, Switzerland\\
C. Marin Benito, Universit{\'e} Paris-Saclay, CNRS/IN2P3, IJCLab, Orsay, France \\
A. Mathad, Universit\"at Z\"urich, Winterthurerstrasse  190, Z\"urich, Switzerland\\
J. McFayden, CERN, Esplanade des Particules 1, Geneva, Switzerland\\
P. Milenovic, University of Belgrade, Studentski trg 12, Belgrade, Serbia\\
V. Milosevic, Imperial College London, Blackett Laboratory, Prince Consort Rd, London, United Kingdom\\
D. S. Mitzel, CERN, Esplanade des Particules 1, Geneva, Switzerland\\
Z. Moravcov\'a, Niels Bohr Institute, University of Copenhagen, Blegdamsvej 17, Copenhagen, Denmark\\
L. Moureaux, ULB, Boulevard du Triomphe, 2, Brussels, Belgium\\
G. A. Mullier, Lund University, Professorsgatan 1, Lund, Sweden\\
M. E. Nelson, Stockholm University and The Oskar Klein Centre for Cosmoparticle Physics, Roslagstullsbacken 21, Stockholm, Sweden\\
J. Ngadiuba, CERN, Esplanade des Particules 1, Geneva, Switzerland\\
N. Nikiforou, University of Texas at Austin, 2515 Speedway, Austin, TX, United States\\
M. W. O\'Keefe, University of Liverpool,  The Oliver Lodge Laboratory, Liverpool, United Kingdom\\
R. Pedro, LIP, Avenida Professor Gama Pinto 2, Lisbon, Portugal\\
J. Pekkanen, University at Buffalo, 239 Fronczak Hall,  Buffalo, United States\\
M.  Queitsch-Maitland, CERN, Esplanade des Particules 1, Geneva, Switzerland\\
M. P. L. P. Ramos, LIP, Campus de Gualtar, Braga, Portugal\\
C. {\O}. Rasmussen, CERN, Esplanade des Particules 1, Geneva, Switzerland\\
J. Rembser, CNRS/IN2P3, Institut Polytechnique de Paris, Ecole  Polytechnique, Palaiseau, France\\
E. T. J. Reynolds, University of Birmingham, Edgbaston, Birmingham, United Kingdom\\
M. Sas, Nikhef/Utrecht University, Science Park 105, Amsterdam, Netherlands\\
R. Sch\"ofbeck, HEPHY, Nikolsdorfergasse 18, Vienna, Austria\\
M. Schenk, EPFL, Route Cantonale, Lausanne, Switzerland\\
P. Schwendimann, Paul Scherrer Institute, Forschungsstrasse 111, Villigen PSI, Switzerland\\
K. Shchelina, LIP, Av. Prof. Gama Pinto 2, Lisbon, Portugal\\
M. Shopova, Institute for Nuclear Research and Nuclear  Energy (INRNE) - BAS, blvd. Tzarigradsko Shaussee 72, Sofia, Bulgaria\\
S.  Sekmen, Kyungpook National University, 80 Daehak-ro Buk-gu, Daegu, Republic of Korea\\
S. Spannagel, DESY, Notkestr. 85, Hamburg, Germany\\
I. A. Sputowska, The Henryk Niewodniczański Institute of Nuclear Physics Polish Academy of Sciences, ul. Radzikowskiego 152, Kraków, Poland\\
R. Staszewski, Institute of Nuclear Physics Polish Academy of Sciences, Radzikowskiego 152, Cracow, Poland\\
P. Sznajder, National Centre for Nuclear Research (NCBJ), Pasteura 7, Warsaw, Poland\\
A. Takacs, University of Bergen, Allegt. 55, Bergen, Norway\\
V. T. Tenorth, Max-Planck-Institut f{\"u}r Kernphysik, Saupfercheckweg 1, Heidelberg, Germany\\
L. Thomas, ULB, Boulevard du Triomphe 2, Bruxelles, Belgium\\
R. Torre, CERN, Esplanade des Particules 1, Geneva, Switzerland\\
F. Trovato, University of Sussex, Science Park Rd, Falmer, Brighton BN1 9RQ, Brighton, UK\\
M. Valente, Universit{\'e} de Gen{\`e}ve, Quai Ernest-Ansermet 24, Gen{\`e}ve, Switzerland\\
H. Van Haevermaet, University Of Antwerp, Groenenborgerlaan 171, Antwerpen, Belgium\\
J. Vanek, Nuclear Physics Institute, Czech Academy of Sciences, Husinec - Rez 130, Rez, Czech Republic\\
M. Verstraeten, Universiteit Antwerpen, Prinsstraat 13, Antwerp, Belgium\\
P. Verwilligen, INFN sezione di Bari, Via E. Orabona 4, Bari, Italy\\
M. Verzetti, CERN, Esplanade des Particules 1, Geneva, Switzerland\\
V. Vislavicius, University of Copenhagen, Blegdamsvej 17, Copenhagen, Denmark\\
B. Vormwald, University of Hamburg,  Luruper Chaussee 149, Hamburg, Germany\\
E. Vourliotis, National and Kapodistrian University of Athens, Panepistimiopolis, Athens, Greece\\
J. Walder, Lancaster University, Physics Avenue, Lancaster, United Kingdom\\
C. Wiglesworth, University of Copenhagen, Blegdamsvej 17, Copenhagen, Denmark\\
S. L. Williams, University of Cambridge, JJ Thomson Avenue, Cambridge, United Kingdom\\
A. Zaborowska, CERN, Esplanade des Particules 1, Geneva, Switzerland\\
D. Zanzi, CERN, Esplanade des Particules 1, Geneva, Switzerland\\
L. Zivkovic, Institute of Physics Belgrade, Pregrevica 118, Belgrade, Serbia
}

\newpage
\section*{\abstractname}

\normalsize

A group of Early-Career Researchers (ECRs) has been given a mandate from the European Committee for Future Accelerators (ECFA) to debate the topics of the current European Strategy Update (ESU) for Particle Physics and to summarise the outcome in a brief document~\cite{newsletter3}. 
A full-day debate with 180 delegates was held at CERN, followed by a survey collecting quantitative input. 
During the debate, the ECRs discussed future colliders in terms of the physics prospects, their implications for accelerator and detector technology as well as computing and software. 
The discussion was organised into several topic areas. 
From these areas two common themes were particularly highlighted by the ECRs: sociological and human aspects; and issues of the environmental impact and sustainability of our research.
The following paragraphs summarise the key outcomes.

\paragraph{General} The ECRs feel that the attractiveness of our field is at risk and that dedicated actions need to be taken to safeguard its future. 
When continuing on the current path, the field will likely be unable to attract the brightest minds to particle physics. 
The ESU must therefore include sociological and sustainability aspects in addition to technical ones related to machine feasibility and particle physics research.
It is of high priority that funding for non-permanent positions is converted to funding for permanent positions, i.e.\ fewer post-docs in exchange for more staff.
In addition, particle physics should play an exemplary role for sustainable behaviour, being inspirational for both society and other sciences.
Overwhelming consensus was reached on the idea to establish a permanent ECR committee as part of ECFA. 
Such a committee would be able to give a mandate to a few individuals representing the ECRs in various bodies. 

\paragraph{Future of the Field} Among the many open questions in particle physics, the ECRs find Dark Matter, Electroweak Symmetry Breaking and Neutrino Physics to be the three most important ones. 
While being open for future international projects, the ECRs emphasise the importance of a European collider project soon after HL-LHC. 
Postponing the choice of the next collider project at CERN to the 2030s has the potential to negatively impact the future of the field.


\paragraph{Comments on the Briefing Book} The importance of understanding the Higgs mechanism is already well underlined in the Physics Briefing Book (\BB)~\cite{BB}~and shared by the ECRs.
Many ECRs stated their discomfort about the way the full CLIC and FCC programmes were compared, especially by how the different states of maturity of the projects were not taken into account sufficiently. 
Additionally, the impact of collider projects outside Europe on the straw-man scenarios and therefore on the future of the European particle physics landscape has not been laid out sufficiently.

\paragraph{Human and Sociological Factors} For the ESU to be effective and sustainable, it is imperative to holistically include social and human factors when planning the future of the field. 
Therefore, the ECRs strongly recommend that future project evaluations and strategy updates include the social impact of their implementation.
Specifically, equal recognition and career paths for the various domains of our field have to be established to maintain expertise in the field. 
The possibility for a healthy work-life balance and the reconciliation of family and a scientific career is a must. 

\paragraph{Environment and Sustainability} Large European organisations and laboratories such as CERN have a unique position and responsibility in society. 
A strong statement from CERN putting the environment and sustainability at the forefront of decision-making, aiming at becoming a carbon-neutral laboratory in the short term future, would have a significant impact. 
The energy efficiency of equipment and the power consumption of the future collider scenarios are already considered but this should be extended to building insulation and the environmental impact of construction and disposal of large infrastructure.
There should be further discussion of nuclear versus renewable energy usage and CERN could and should strive for a higher renewable energy fraction.
More considerations should be put on the impact of computing and software resources.
Travel and conference schedules should be seriously assessed to reduce the amount of travel and the associated carbon footprint.

\paragraph{Accelerator and Detector R\&D}  Among the ECRs, 88\% are in favour of an $e^+e^-$ machine as the next collider to be realised.
A strong and diverse R\&D programme on both accelerators and detectors must be a high priority for the future.
On the accelerator side, concerns have been raised about whether the key numbers stated in the \BB~allow for a fair comparison of the various projects; while concerning the detector side, there was no mention of which technology is suitable for which future project and the level of readiness of each technology.

\paragraph{Computing and Software} Software and computing activities must be recognised not only as  means to do physics analyses, but as research that requires a high level of skill.
Innovation in physics analysis should strive to minimise the time to produce physics results allowing more person-power to be allocated to areas where innovation and development is truly needed.
In an effort towards reducing the carbon footprint associated with travel for work purposes, our community can drive the development of new software for remote meetings, even for large groups of attendants.
Furthermore, the researchers are generally in favor of open data and see the need for sharing knowledge and resources with other computing communities.

\paragraph{Electroweak and Strong Interaction Physics}
Due to the different demands of electroweak and strong physics, there was no clear consensus in the electroweak and strong interaction physics discussion session as to which future collider should be pursued at this stage. 
For most Higgs couplings, decay width and electroweak precision measurements, $e^{+}e^{-}$ colliders have a clear advantage. 
However, this is balanced by the greater precision that proton-proton colliders have for rare Higgs couplings, in particular the Higgs self-coupling. 
It was noted that $e^{+}e^{-}$ colliders are very appealing due to the shorter timescales and their capability of running at the precise energies required to produce copious amounts of Higgs, $W$ and $Z$ bosons or top quark pairs.
 This is in contrast to the strong physics discussions, where a clear preference towards a $pp$ or $ep$ collider was voiced. Priority should be put on precision measurements and global fits rather than model-driven searches. Tighter collaboration between theory and experiment would enhance the precision of measurements.

\paragraph{Beyond Standard Model, Dark Matter and Dark Sector Physics} No clear consensus on future collider scenarios was found as different, equally valid, theoretical models can prefer one scenario over the other. 
However, the ECRs consider the diversification of experiments, building on projects such as Physics Beyond Colliders, as vital for the future of the field and should be pursued with high priority in parallel to the larger projects. Similarly to the discussion on electroweak and strong interaction physics, it is felt that better collaboration between theory and experiment is needed to extract the full potential of future programmes, more focus on this for ECRs (i.e. including more PhD students) is needed in particular.

\paragraph{Flavour, Neutrino and Cosmic Messenger Physics}
While the heavy flavour domain benefits from any future collider, numerous specialised smaller experiments in the light sector are needed outside these large-scale scenarios to complete the picture.
Real-time observations between connected observatories, for example neutrino, gravitational wave and gamma ray telescopes, will be crucial in the future, and to fully realise the potential in this area synergies with HEP would be vital.

\newpage
\section{Introduction}
This report aims to provide input from the ECR community on the ESU scheduled to be approved by the CERN Council in 2020.

\begin{figure}[!th]
\begin{center}
    \includegraphics[width=0.95\textwidth]{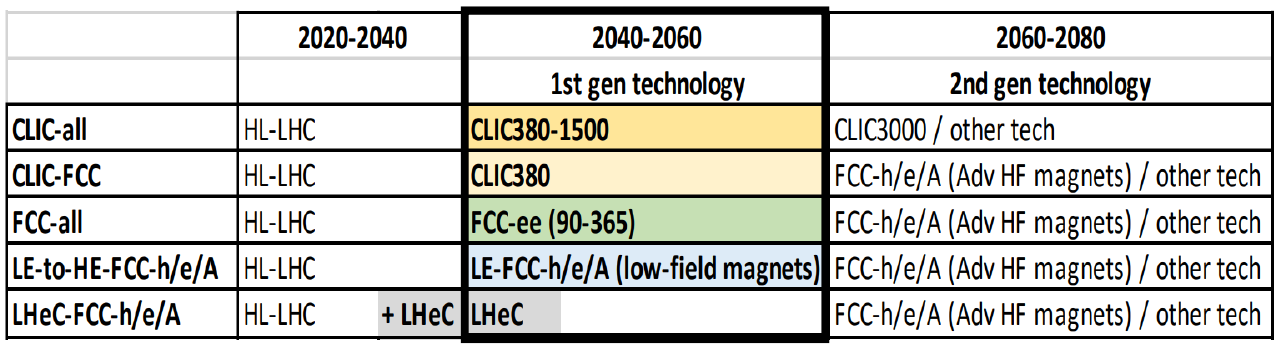}
    \includegraphics[width=0.95\textwidth]{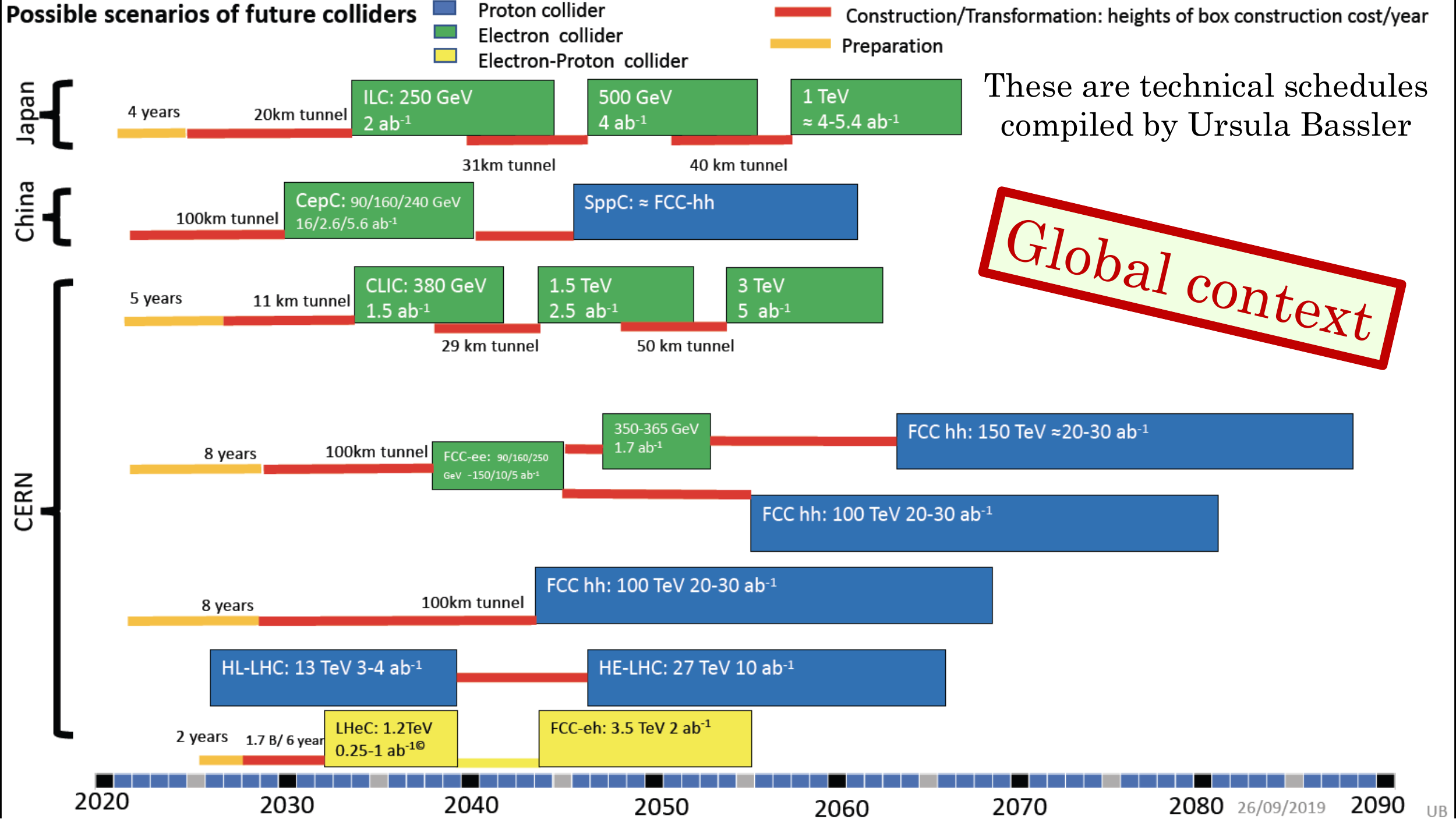}
    \end{center}
    \caption{Five possible future collider scenarios considered for CERN (top) and possible scenarios of future colliders in a more global context (bottom). Slides taken from slides by Prof. Jorgen D'Hondt at the ECR debate on November 15th, 2019.}
    \label{fig:strawManScenarios}
\end{figure}

This initiative was started towards the end of the consultation period for the ESU, which has taken place throughout 2019. 
A total of 180 ECRs from institutes across Europe were invited to a plenary debate on 15th November 2019. 
The broad range of possible topics was subdivided into several areas with common physics or themes to streamline the discussion:
\begin{itemize}
    \item Environment and Sustainability,
    \item Electroweak and Strong Interaction Physics,
    \item Beyond-the-Standard Model (BSM), Dark Matter and Dark Sector,
    \item Neutrino, Flavour and Cosmic Messenger Physics,
    \item Accelerator and Detector R\&D, and
    \item Computing and Software.
\end{itemize}
At least one meeting was held within each working group in the weeks before the plenary debate. 
The framework for the discussions in the meetings was given by the 2019 Physics Briefing Book (\BB) and the different straw-man scenarios \cite{CERN-ESU-005} currently under consideration for the future of CERN (Figure~\ref{fig:strawManScenarios}). 
The working groups were also tasked to identify issues that are important to the ECRs but not covered in the Briefing Book.
Each working group presented the outcome of their meetings in the plenary session, followed by a discussion.
The organising committee, a self-nominated subgroup of the ECRs, noted that the subject of human and social factors overarches all working groups, so it was allocated a dedicated session in the plenary debate.

For the ECR debate, each of the working groups and the \textit{human and social factors} group summarised the main points discussed in the meetings prior to the debate and each summary was followed by a discussion. 
Additional time was allocated at the end of the debate to allow further discussion on topics not covered during the previous sessions.
To obtain some quantitative information about the views of the delegates the organisation committee also prepared an online survey for all delegates after the debate. 
The results of this survey complement the summary of the plenary debate and are also included in this report.

\section{Survey Results}
\label{sec:survey}
A survey was circulated to all the ECRs nominated by ECFA. 
In this section an overview of the results is presented. 
A full breakdown of the answers is listed in the Appendix~\ref{appendix:socialsurvey}.
There was not yet time to perform a more detailed analysis of responses including correlation between or isolation of different demographic groups.

\subsection{Demographics}
\begin{figure}
    \begin{center}
    \includegraphics[width=0.48\textwidth]{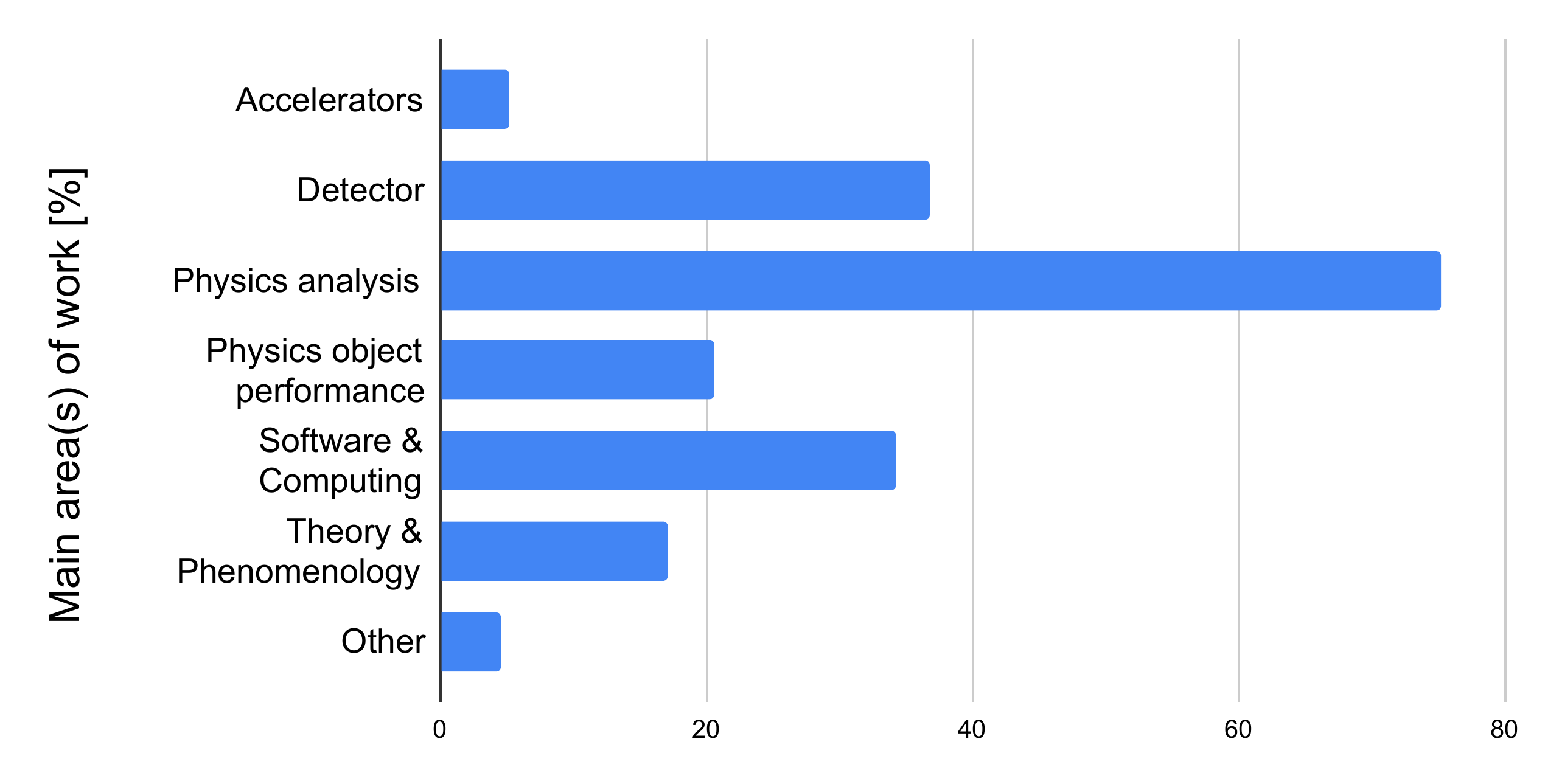}
    \includegraphics[width=0.48\textwidth]{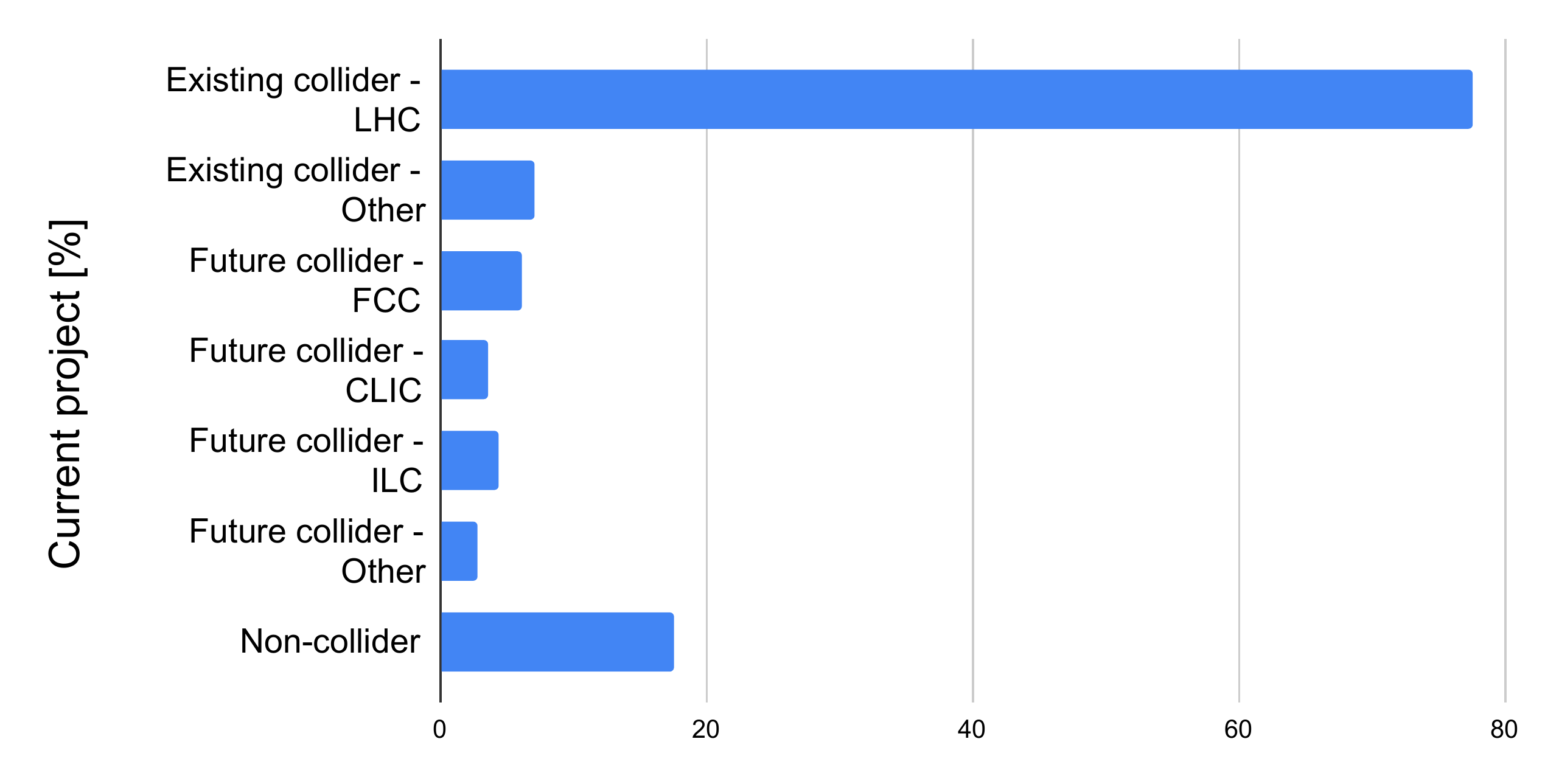}
    \end{center}
    \caption{Current area(s) of work (left) and projects (right) of the ECRs.}
    \label{fig:survey_currentwork}
\end{figure}

There were 117 responses to the survey, corresponding to almost two thirds of the 180 invited participants.
Responses came from people of many different nationalities, spread across a wide range of locations in Europe.
Over 80\% of participants are either PhD students (33\%) or post-docs (50\%). Two thirds of people were less than 7 years past the start of their PhDs and 88\% less than 10 years. 
The gender of people responding to the survey was 63\% male and 33\% female, with 4\% preferring not to specify.
The current areas of work and projects are summarised in Figure~\ref{fig:survey_currentwork} with about three quarters of the ECRs being mainly involved in physics analysis and a similar fraction performing work related to the LHC or HL-LHC. 


\begin{figure}
    \begin{center}
    \includegraphics[width=0.48\textwidth]{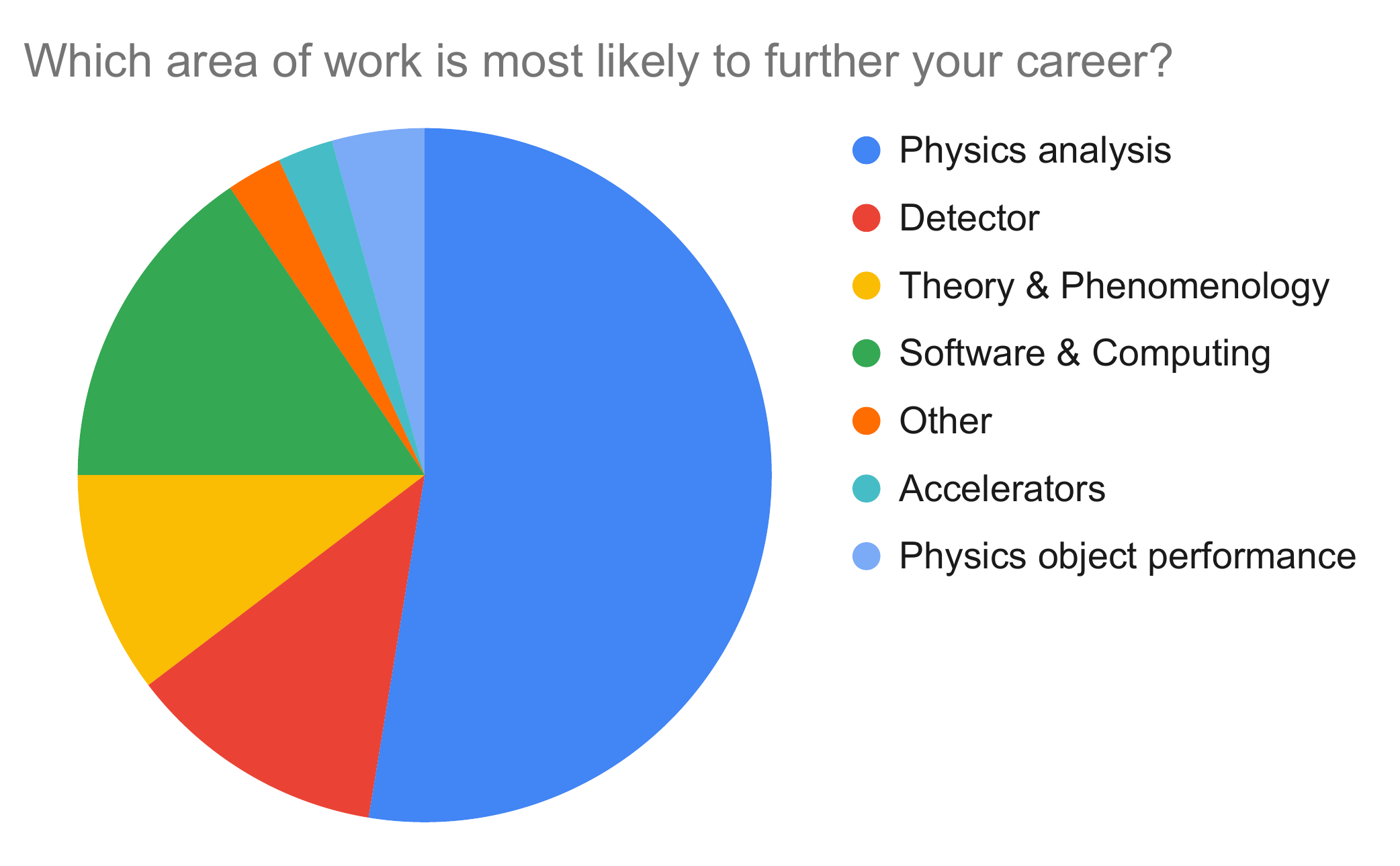}
    \includegraphics[width=0.48\textwidth]{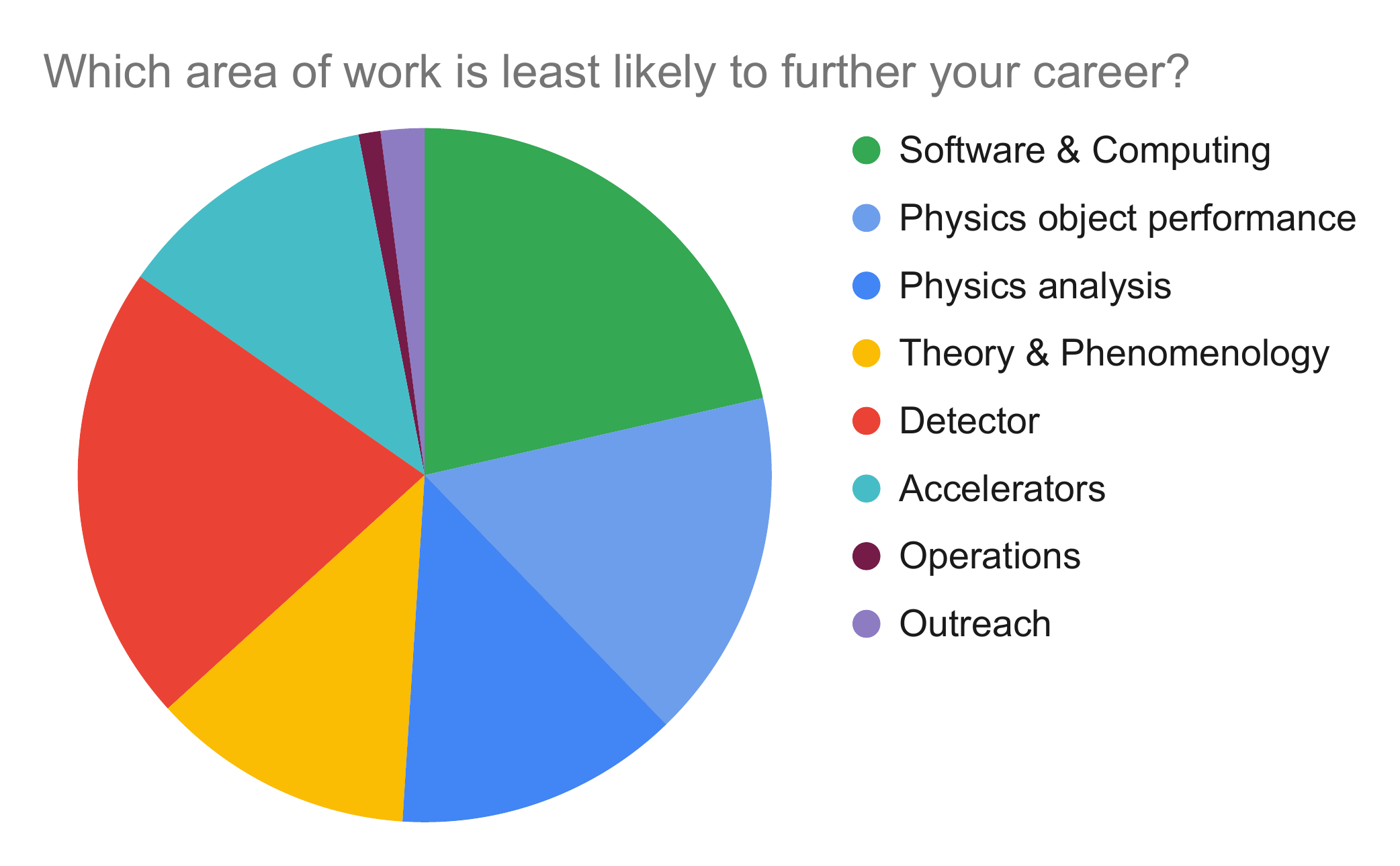}
    \end{center}
    \caption{The areas of work which ECRs think are most (left) and least (right) likely to further their careers.}
    \label{fig:survey_areasforcareer}
\end{figure}

\subsection{European Strategy Update}
Over 60\% of people felt well informed about the ESU process and topics and around 85\% felt that the ECR discussion process (meetings prior to the debate, documentation provided and the debate itself) improved their knowledge of the ESU. 

\begin{figure}
    \begin{center}
    \includegraphics[width=0.7\textwidth]{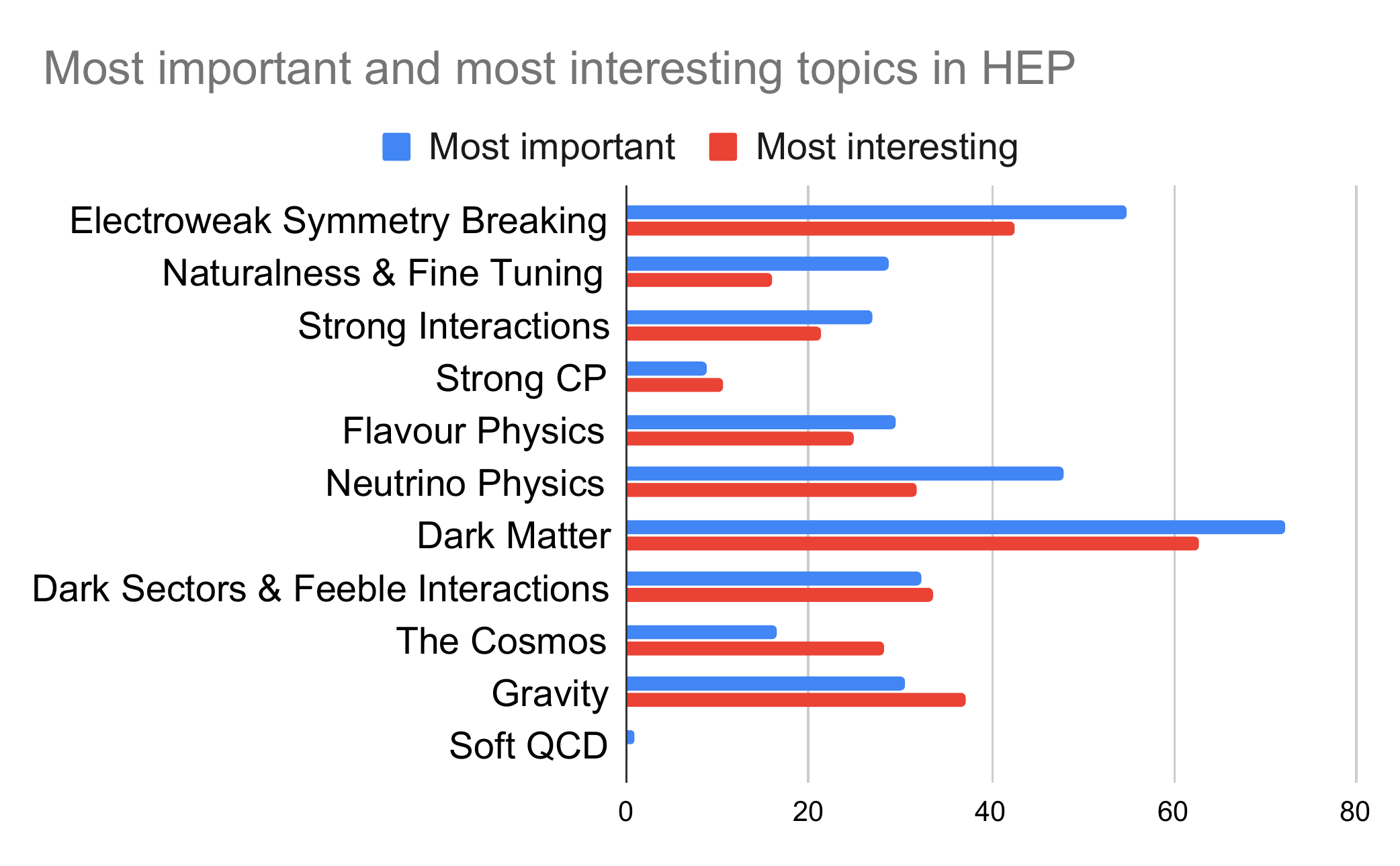}
    \end{center}
    \caption{Most important and most interesting topics in HEP.}
    \label{fig:survey_importantinteresting}
\end{figure}

The open topics/questions in HEP (as defined in the \BB) that ECRs consider most important and most interesting are shown in Figure~\ref{fig:survey_importantinteresting}. 
Dark Matter is clearly ranked highest in both most interesting and most important categories. 
Electroweak Symmetry Breaking is second, but by a small margin. Neutrino Physics is ranked third in importance and Gravity is ranked as the fourth most interesting. Overall there is quite a diverse distribution of responses, that well reflects the diversity of the field.

\begin{figure}
    \begin{center}
    \includegraphics[width=0.47\textwidth]{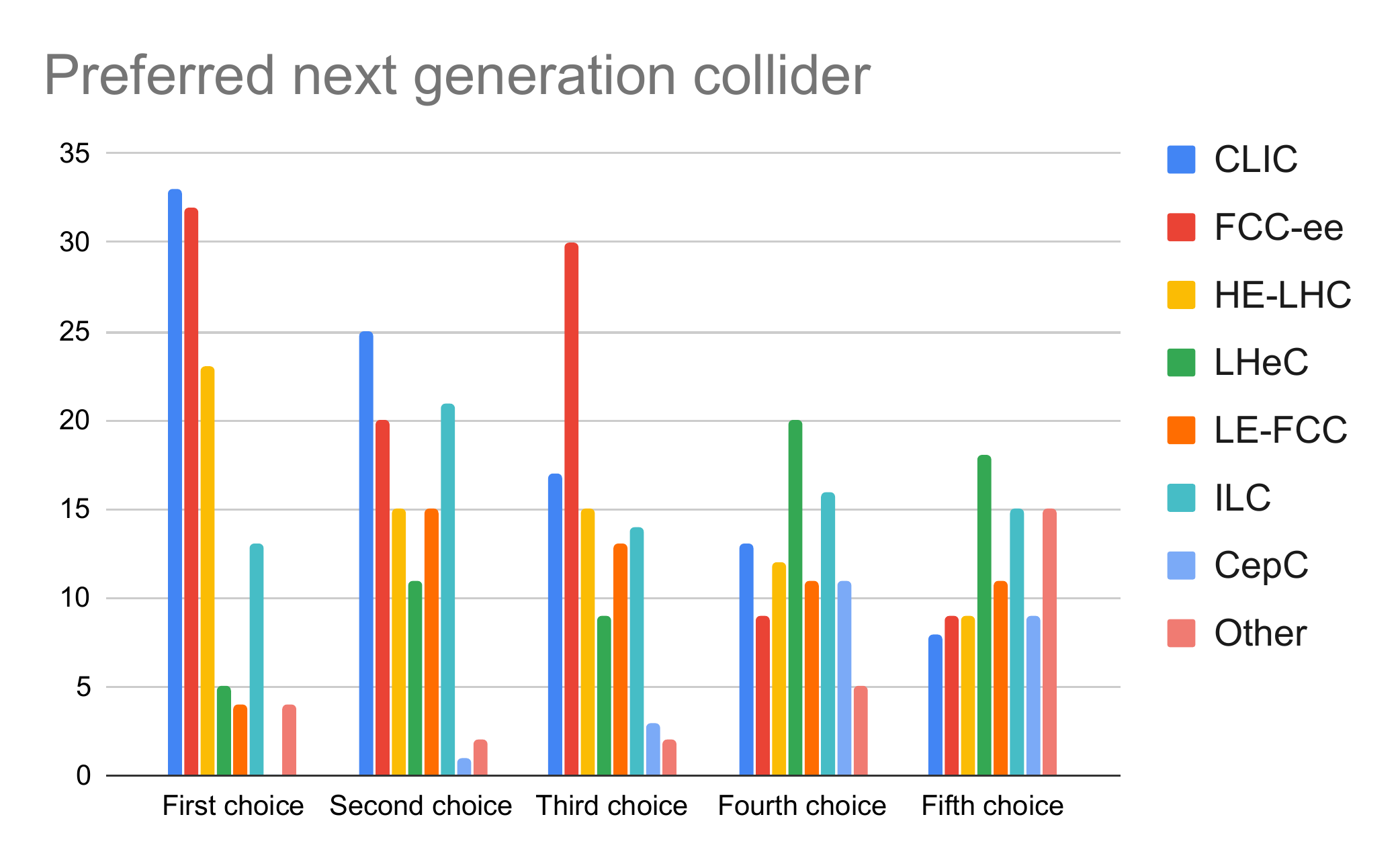}
    \includegraphics[width=0.49\textwidth]{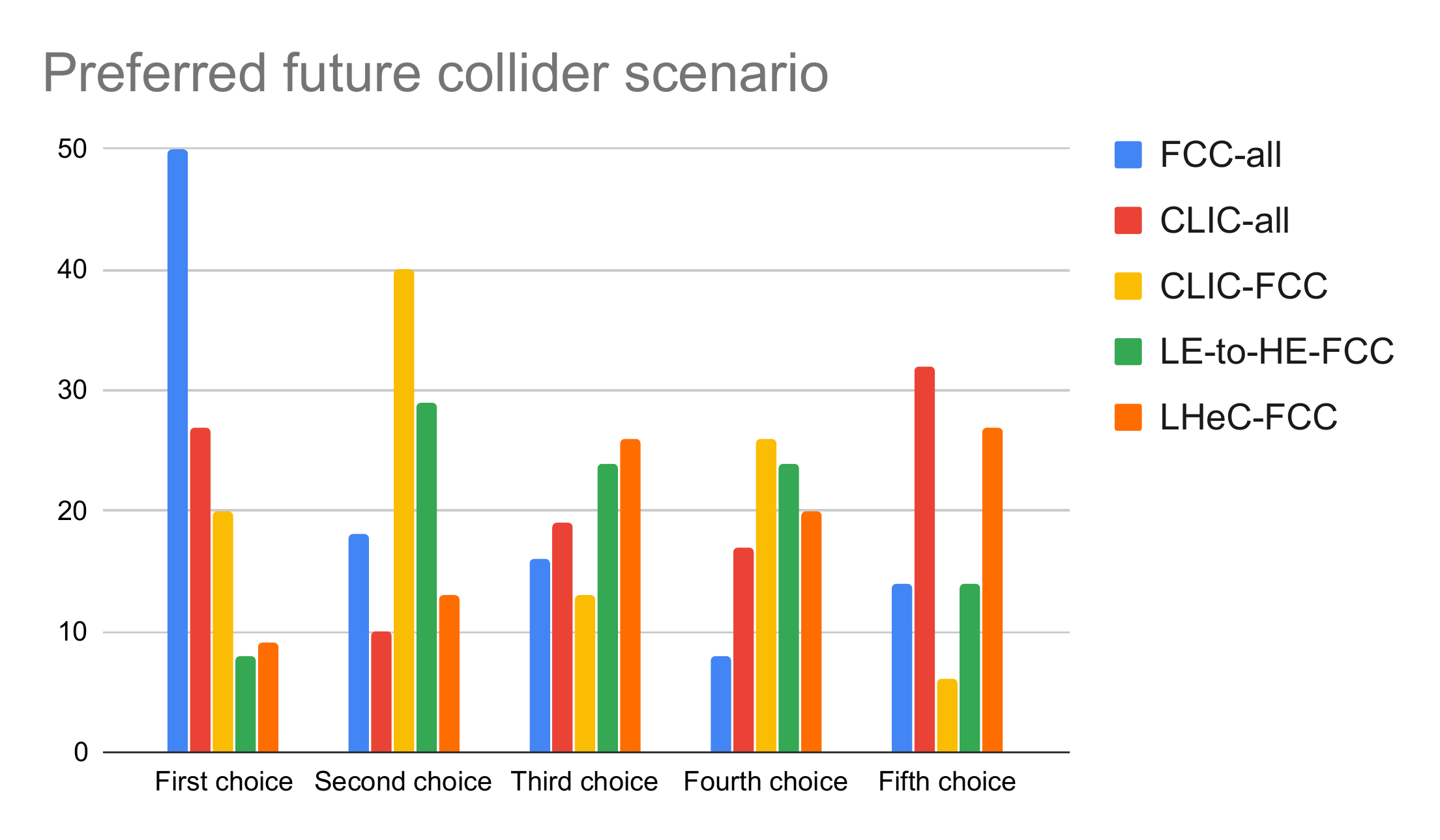}
    \end{center}
    \caption{Left: Preferred next-generation collider. Right: Preferred future collider scenario at CERN.}
    \label{fig:survey_nextgen_future}
\end{figure}

ECRs were asked how important they think it is that Europe should build a large collider after HL-LHC. Over 85\% of people felt that this was important.
70\% of ECRs agreed that the next-generation collider should be an $e^+e^-$ machine. 
The survey asked which next-generation collider and which of the long-term future collider scenarios (as laid out in Figure~\ref{fig:strawManScenarios})  the ECRs would prefer. 
The responses are shown in Figure~\ref{fig:survey_nextgen_future}. CLIC and FCC-ee seem to have similar levels of support as the next-generation collider. 
If the next $e^+e^-$ collider is built in Asia, the preference of ECRs is clearly for CERN to embark on the full FCC programme. 
Of the FCC projects, the largest interest was in FCC-hh followed by FCC-ee, which later transitions into FCC-eh.

For the long-term future of the field there is an overall preference for the full FCC scenario. 
The second favourite choice is the full CLIC program. 
There is also significant support for CLIC380 followed by FCC. 

When asked explicitly whether CERN should build CLIC and FCC, 65\% of participants agreed that FCC should be built compared to 41\% that thought CLIC should be built. 
84\% of people agreed that not building a collider at CERN soon after HL-LHC will negatively affect European particle physics research.

\subsection{Sociological and Human Aspects}

Several questions on the survey focused on sociological and human aspects related to the ESU. 
90\% of the respondents indicated that they are interested in a long term career in research. 
Anxiety associated with their career prospects was felt by
71.2\% of respondents, with only 35.1\% indicating agreement that they felt they had good prospects for a long-term academic career. 
Only 6\% of respondents disagreed with the statement that they had the skills necessary for a non-academic career. However, the proportion indicating that they felt they would be given the necessary support in finding a non-academic job was only 22.4\%.
Concerning a career in particle physics, about 50\% of the ECRs stated that physics analysis is the area of work that would most likely further their career, as is shown in Figure~\ref{fig:survey_areasforcareer}.
Similarly, about 40\% think that detector-related work and software and computing related work are least likely to further their careers. 

Regarding working hours and conditions, 77.6\% of respondents either agreed or strongly agreed that working overtime was a necessary ingredient to secure an academic career. 
70.3\% of participants responded 4 or 5 when rating their levels of work-related stress, with 45\% responding in the same way to experiencing work-induced stress burnouts. 
To the question concerning discrimination at work due to gender/race/sexuality or other factors, 
almost 20\% replied with (4) or (5), indicating the need for substantial improvement in this area.

Two of the survey questions were posed to the ECRs regarding the time spent on job applications and doing pure research.
76.9\% of ECRs responded that they spent less than 20\% of their time applying for jobs/grants. However, only 9.3\% indicated they spent $>$80\% of their time on pure research, with 66.9\% having between 40-80\% of their time for research.

Challenges associated with balancing an academic career with family life were discussed as part of the debate. 
Of the participants in the survey, about 80\% do not have children, with 70\% of participants indicating an agreement with the statement \textit{``having children would negatively impact my academic career''}. 
55.7\% of respondents agreed with the statement that \textit{``I feel at ease in expressing concerns about reconciliation of work and personal/family life in the workplace''}. 

A common experience for ECRs is the need to move countries as a result of their work. 
Among the ECRs, 81.8\% indicated either agreement or strong agreement with the statement \textit{``If you had to change your country during your career, do you think it has had a positive impact on your career?''}, however 66.4\% indicated disagreement with the statement that \textit{``If you had to change your country during your career, do you think it has had a positive impact on your personal/family life?''}. 
Two questions concerned the possibility of remote working: 53.4\% of respondents indicated agreement with the statement \textit{``My job could be performed effectively working remotely''} with a similar number (50\%) agreeing that \textit{``The ability to work remotely more often would increase the appeal of a particle physics career''}. 
However the number of neutral responses for these questions was higher than most others in this section, at 23.7\% and 28.4\%, respectively.

\subsection{Environmental and Sustainability Considerations}

The environmental concerns of ECRs are illustrated by the fact that 97\% of people felt that environmental impacts should be taken into account when taking decisions on future projects, with 40\% considering this an \textit{extremely} important factor.
Attending conferences in person was considered to be necessary to secure an academic career by 87\% of the participants, with 75\% of people agreeing that not attending conferences because of environmental concerns would harm their career. 
However, 78\% of people would attend conferences remotely more often if better tools were available.

Just over half of the participants feel at ease expressing their environmental concerns at the workplace. 
The same fraction of people would be willing to ask their employer to purchase carbon offsets for their work-related flights. 
Exactly half would be willing to pay themselves.

\subsection{Other questions}

92\% of the respondents were in favour of opening data to the public. 
Asked what the main purpose of this open data should be, three use cases were the most popular: scientific goals, education and outreach. 
76\% of the people agreed that this data should have an embargo periodo of a few years before being released.

On the topic of recognition for different areas of work, only 37\% of the people felt that detector R\&D and software and computing work is well-recognised in their experiment or group. 
53\% of the people thought that detector R\&D and software and computing work are as important as physics analysis for their future career in HEP.

\section{Summary of Discussions During the ECFA ECR Debate and Working Group Meetings}

\subsection{Sociological and Human Aspects}

This section summarises the ECR discussion on social and human aspects related to a research career in particle physics in general and to future long-term projects in particular. 
Such topics were only partially discussed in the \BB~and the ECRs would like to emphasise their importance in this report. 
For the ESU to be effective and sustainable, it is imperative to holistically include social and human factors. 
In the future, a metric measuring the success of the particle physics community should also include improvements in diversity, reconciliation of family and scientific career and health of the individuals.
Furthermore, the ECRs strongly recommend future project evaluations and strategy updates to include the social impact in their implementation.
Besides CERN, universities and national laboratories are encouraged to implement such improvements and measures. 
Therefore, a significant fraction of project-oriented, short-term funding has to be converted into base-funding, enabling the institutions to realise a healthier and more family-friendly environment with a larger fraction of permanent positions.

\subsubsection{Future Career Opportunities}
According to the survey, most of the ECRs would like to continue to work in the field, but they are worried about their career prospects in academia.
Researchers aiming to stay in the field are often required to move from one institution to another, often involving a change of country. 
About 80\% of the ECRs recognise that this has positive impact on the future career and enriches the scientific background but also represents a major challenge for personal aspects. 
As a result, some ECRs feel that the mobility-policy of many grants discriminate people with families, and therefore they request relaxing the importance of mobility for such applications, given the possibility of remote collaboration.
Concerning large experiments outside Europe, many ECRs see challenges associated with travelling to these experiments, which is partially, but not exclusively, due to family reasons and is considered less of an issue for researches already working in experiments based in Asia or in North America.

Future colliders together with their experiments within and outside of Europe will offer many opportunities in the field for young researchers.
Almost all ECRs agree on the importance of a European collider project after HL-LHC, but severe concerns about the lack of risk assessment related to the timescales and possible delays of the projects have been raised. 
Thus, CERN should carefully plan its strategy to maintain its leading role in the European particle physics landscape and strive to serve as a hub for any such international projects outside Europe.
In order to successfully implement large-scale projects, it is of highest priority that expertise in all domains is kept over the entire preparation phase and beyond.
As a community, we cannot afford to lose a large fraction of experts within certain phases, and expertise has to be continuously passed on to younger individuals.
However, a prompt continuation of a new collider after the end of the HL-LHC might be less likely for very large projects, posing a risk in losing expertise and attractiveness of the field.

As for career possibilities outside HEP, more than 75\% of the ECRs think that they have the skills to pursue a successful career in a non-academic job. 
Ability to collaborate in large software infrastructures, experience with big data, management and problem-solving skills gained through scientific work are sought-after skills that are recognised in non-academic sectors. 
However, only 22.4\% of the ECRs think that they would be given the support needed to find a suitable career outside HEP. 
In this respect, alumni networks within Europe could be helpful and thus some ECRs suggest to introduce more of such networks and to strengthen existing ones. 
Moreover, the possibility of contributing from outside HEP or returning to particle physics after a non-academic job can be very beneficial for the field but presently this is difficult to realize.

\subsubsection{Towards Equal Recognition}

As also stated in the ECFA Detector Panel Report~\cite{surveyDetector}, detector and software technology research are often less valued than physics data analysis and interpretation, although they play an essential role for progress in experimental particle physics.
Tasks related to this research are, unfortunately, too often recognised only as a means to extract data from the experiments, while they should be considered proper research areas in their own right. 
This can be particularly dangerous to our field because it reduces the number of high-level creative professionals with the specialised expertise needed in experiments with high complexity.
An equal recognition of tasks should be encouraged within experiments, universities and laboratories in order to improve the current situation. 
The ECR delegates propose the following actions: 
\begin{itemize}
    \item Establishing professional career opportunities for all domains of expertise, for example through the creation of advanced research grants and long-term positions up to the professorship;
    \item Devoted awards inside the collaborations in the domains of detector/accelerator R\&D, computing and analysis should be introduced with the aim of balancing recognition and increasing the visibility of individuals. Moreover, awards should be considered also for researchers working on a mixture of the above topics, which is a common case;
    \item More publications on R\&D topics, especially on software and computing works, should be encouraged to increase recognition outside the collaborations.
\end{itemize}

\subsubsection{Reconciliation of Family and Scientific Career}

Reconciliation of family and a scientific career is imperative, but is far from being realized. 
About 70\% of the ECRs feel that having children would negatively affect their academic career in the research field. This problem is particularly relevant to ECRs because the age at which they usually decide to have children may correspond to the period in which they are still aiming for a permanent position.  This percentage goes up to 85\% in the case of women. This issue, together with historical inheritance, can be easily connected to the drop out of women in physics especially in the first stages of the career, which is still a common issue in European countries. Encouragement of longer and shared parental leaves would help all researchers who feel that starting a family is a threat to their careers.

It is important to mention that reasonable work-life balance should not come into effect only after a successful permanent job application, but should be optimally achieved also in the first stage of the scientific career.
More than half of the ECRs think that their job could be performed remotely. In general, more possibilities of remote work would increase the appeal of a particle physics career and engage more researchers with families as conveners or in committee work, which would be beneficial for their career. 
Moreover, having children at any stage of the scientific career often limits the possibilities to take part in conferences and workshops.
The importance of being physically present in meetings and conferences to increase visibility and to improve the career prospects was seen by more than 85\% of all ECRs. 
However, during the plenary debate the importance of a more sustainable and family-friendly approach was discussed.
Ideas such as increasing assistance provided by local organisation committees for researchers travelling with children and to improve remote participation with better video-conferencing software were very much endorsed, as also discussed in Section~\ref{sec:env}.

Today's generation aims for a healthy work-life balance, with or without children. 
Although working extra hours is considered acceptable when required, about 70\% of the survey participants feel stressed and 45\% reported to have experienced burnouts related to work.
This situation is not acceptable and reflects the severe risk of (self-)exploitation in the community.

\subsubsection{Diversity}
In the opinion of the ECR delegates, some groups are still under-represented both inside and outside academic fields, with different phenomenology and history in each country. 
The authors strongly support and promote inclusion, especially in an increasingly global community.
Therefore, we suggest the establishment of diversity officers in each collaboration/facility/institute, to whom diversity related issues can be reported. 
This is already in place in some collaboration and should be encouraged.

\subsection{Environment and Sustainability}
\label{sec:env}

The ECR community believes that, as scientists, it is our responsibility to play an exemplary role in tackling humanity's challenge of climate change.
In order to have a chance to prevent Earth's average temperatures from rising by more than 1.5$^{\circ}$ C compared to the pre-industrial era,  emissions of greenhouse gases (GHGs) need to be drastically reduced worldwide, across all sectors, on a timescale of ten years, and eventually be driven to net-zero by 2050.

Particle physics research has always been at the forefront of science and technology, exploring uncharted territories and devising the tools and methods that made it possible. Technological challenges such as analysing previously unheard-of amounts of data, or detecting extremely rare events in a highly radioactive environment, have been successfully tackled. Particle physics research is a driver for technological innovation, and its impact on society and economy is significant.

CERN and other major European laboratories must have an ambitious vision and develop clear action plans to become carbon neutral. 
As a leading research centre, CERN's vision should be inspirational for both society and the scientific community, and it would lead to dramatic advancements in energy-saving techniques, renewable energy sources, and public awareness.
We believe such a commitment would eventually lead to an increased budget for physics research thanks to a better consideration of research activities by the general public, eventually favoring important political decisions.
The ECR community is ready to embrace the radical shift of paradigm that is needed in order to make our life and research work sustainable, carbon neutral, and exemplary for the changes we wish societies will undertake.

\subsubsection{Comments on the Briefing Book}
The \BB~addresses the topic of sustainability in Chapter 10, ``Accelerator Science and Technology'', specifically in the few paragraphs of Section 10.7, ``Energy management''. 
We find these paragraphs to be lacking and to not contain sufficiently precise recommendations.
Besides, while so far attention has only been focused on the energy efficiency of HEP equipment, detailed studies should also be carried out about the environmental impact of the same equipment's construction and disposal.

\subsubsection{Instrumentation, Operation and Computing}
The ECR community hopes that leading research centres such as CERN can negotiate with energy providers in order to ease the switch to renewable energy sources.

CERN's accelerator complex has a yearly power consumption in the TWh range. The CERN data centre remains in the range of about 50~GWh. Computing accounts for a few percent of worldwide CO\textsubscript{2} emissions. 
At CERN, computing is minor with respect to accelerator operation, but several other grid sites operate around the globe. Servers and storage systems consume power and need to be cooled. 
The possibility of relatively easy improvements has been demonstrated for example at the GSI Helmholtzzentrum f\"ur Schwerionenforschung with the Green IT cube~\cite{GSIGreenIT}. 
Moreover solutions invented at CERN can be exemplary for other large computing sites.
The ECR delegates urge the HEP community to rethink our data processing needs in terms of sustainable computing.

Another successful example of resource-usage mitigation is the recycling of waste heat. 
This solution is already in place at the Paul Scherrer Institute (PSI), where roughly 60\% of the heating needs are covered by waste heat recovery, the rest is supplied as remote heat by the nearby nuclear power plant~\cite{PSIWasteHeat}.
The waste heat recovery project was initiated in 2010 with an investment below 4\,MCHF. Break-even is expected in 2021 after 10 years of operation, and it will be followed by yearly savings of the order of 0.5 MCHF thereafter. 
A similar project has been initiated at the LHC Point 8, with an agreement signed by CERN and the French authorities aimed at using waste warm cooling water from the LHCb experiment to heat a new residential area in the city of Ferney-Voltaire
\cite{FerneyVoltair}. 
We suggest to study similar solutions at the locations of the other present LHC experiments (and of those to come), including improved insulation for existing buildings.

Any investment aimed at lowering energy consumption, including heating of buildings, resource usage and computing resources usage,
or aimed at mitigating the environmental impact of instrumentation and facility construction and
disposal will have an immediate or long-term  financial return. 
CERN and other major laboratories will profit from long-term savings after investing in energy efficiency.

\subsubsection{Societal Considerations}
The role of CERN as a science-disseminating centre in Europe and worldwide is clear.
The laboratory’s public image will greatly profit from implementing an ambitious strategy for
sustainability. It might in fact suffer from not doing it.
If CERN and other major laboratories were to invest in research for the application of renewable energy and sustainability, they could potentially become globally renowned role models. 
This would positively affect the rest of the society as well.

We believe that steps should be taken in order to diversify investments, which should explicitly span from physics to sustainability and also engage in improving the scientific education of the general public. We also believe that such diversification would eventually lead to an increased budget for physics research, stemming from positive impact on society, and lead to a better general view of research activities. In addition, positive attitude towards physics from the public could lead to favourable political decisions.

We urge laboratories to consider taking small actions in the very near future. Such actions would have a limited direct impact on the environment, but a broad induced impact from raising awareness among scientists and the general public. A non-comprehensive list of such relatively inexpensive small actions includes:
\begin{itemize}
\item Replacing or reducing climatisation, for example by adding vegetation on lab and office roofs, and reducing the sometimes unnecessary heating of offices;
\item Adding solar panels;
\item Gradually renovating old offices at CERN, installing proper insulation and implementing environmentally-friendly architecture wherever possible;
\item Holding conferences in places reachable by train by most attendees;
\item Reserving budget to offset non-eliminable carbon emissions;
\item Promoting biking and the use of public transport;
\item Promoting the use of electric vehicles (charging from renewable energy sources);
\item Promoting responsible food sources; and
\item Setting up information about sustainable living and research and training events.
\end{itemize}

\subsubsection{Human and Career Considerations}

Concerns have been raised about air travel, which represents about 2\% of global CO\textsubscript{2} emissions\footnote{A single flight from London to Geneva emits on average 400 kg of CO\textsubscript{2} in the atmosphere, per passenger. This figure is equivalent to the carbon absorption of one tree's lifetime. For a carbon-neutral journey, each passenger should arrange for a short-term carbon intake of 400 kg, which can be achieved in about 10 years by planting about 10 trees per passenger.}. Scientists, and in particular ECRs, face significant pressure to travel because of the well-established link between a researcher's visibility and his/her value as scientist.
At the same time, conference software and virtual reality tools are still inadequate and have little support.
This frame of mind leads to an unsustainable and unnecessary number of journeys, and at the same time discriminates against ECRs with family or local commitments.
To make things worse, the job market in academia is highly competitive, leading ECRs to prioritise career concerns over environmental considerations. Some ECRs think that taking up a clear stance for the environment can damage their career prospects.

In LHC experiments, leadership, responsibility and coordination positions such as convenerships are often assigned by taking into account the candidate's presence at CERN. The ECR community doubts that this is necessary in all cases. Long-distance commitments should be enabled and promoted.

\subsubsection{Requests of the ECR Delegates for the ESU}
The ECR community has agreed upon the following list of points to be addressed in the context of the ESU. 
We strongly request that:
%
\begin{itemize}
    \item CERN and the other major research laboratories create a permanent committee in charge of establishing and enforcing sustainability criteria;
    \item New projects and upgrades of existing projects submit a detailed analysis of environmental impact along with the Technical Design Report or equivalent design document;
    \item CERN should seek funding for the transition to carbon neutrality through project proposals targeting European Commission (EC) funds dedicated to climate change;
    \item The climate commitment of countries where projects aim at being hosted be evaluated as part of the project review process;
    \item CERN and the other major laboratories set goals more ambitious than those envisaged by the International Panel on Climate Change (IPCC)~\cite{IPPC}. We strongly recommend to aim at carbon neutrality by 2030. The laboratories should put in place the necessary measures to attract and train professional figures that will make it possible to meet these goals. It is in CERN's interest to take a lead in switching to sustainable, carbon-neutral research;
    \item Global environmental sustainability goals are defined and established as part of this Strategy Update;
    \item The resource intake of particle physics laboratories is transformed, and that the necessary measures are put into place in order to move scientific infrastructure to carbon-free and renewable sources;
    \item Investments are made in order to make better use of the available technologies. The recovery of waste heat is an example already in place at PSI \cite{PSIWasteHeat} and being implemented at the LHC Point 8;
    \item Conference-related and meeting-related air travel is demoted. Videoconferencing should instead be encouraged, and frameworks for remote collaboration improved. When air travel is deemed necessary, carbon offsetting solutions should be considered and made part of the travel budget.

\end{itemize}


\subsection{Electroweak and Strong Interaction Physics}

Electroweak symmetry breaking was voted as the second most important and interesting topic among the ECR delegates. In particular, a better understanding of the Higgs mechanism is imperative for the future of particle physics.
The ECRs discussed the significance of studying the Higgs sector in detail because of its sensitivity to BSM physics, its cosmological consequences and the need of a deeper  understanding of the Higgs sector, as a key part of the Standard Model.
Given the material in the \BB, the best option for measuring Higgs couplings, including the Higgs self-coupling, is to use data from an $e^+e^-$ collider, followed by data from the FCC-hh. 
Electron-positron colliders have the advantage that they allow the tagging of the $ZH$ process, and therefore offer a measurement of the total decay width of the Higgs boson. 
This allows for model-independent measurements of the Higgs bosons couplings, and for the detection of Higgs decays to invisible BSM particles, such as Dark Matter candidates. Lepton colliders also offer unrivaled measurements of the couplings of the Higgs boson to many of the SM fermions.
Furthermore, linear $e^+e^-$ colliders could go beyond the $HH$ threshold and obtain competitive results on the self-coupling measurements.
However, circular lepton colliders can only operate below the $HH$-production threshold, meaning that the only constraints on the Higgs self-coupling would be from single Higgs measurements, or by the application of the next phases of the FCC program (electron-hadron and hadron-hadron).

Regarding BSM searches, it was made clear that higher energies are required for a variety of searches for BSM resonances, strong SUSY and many other scenarios lying at a few TeV. 
While the FCC-hh is currently the highest energy option on the table, a point was made about considering other accelerating technologies, such as plasma wakefield acceleration. 
It was pointed out that there is also the possibility to find new physics resonances at lower energies, for example profiting from high precision measurements at lepton colliders with much reduced backgrounds. 
In addition, the importance of inputs from HL-LHC results to define the goals of future projects in terms of BSM searches was discussed.

Lepton colliders could provide precision measurements of $Z$ and $W$ boson properties, especially if polarised beams and dedicated runs are set in place.
Furthermore, polarised beams would lead to an increase in the number of electroweak observables which can be probed
allowing for the exploration of the full chiral structure of the SM for all fermions (including the top quark, produced far from the $t\bar{t}$ threshold). Moreover, many BSM models predict deviations for the right helicity couplings which will be only measured through polarisation analysis of the inital or final states. Circular colliders such as FCC-ee and CEPC will have larger luminosities than linear colliders below the Higgs threshold, showing superiority in non-polarised observables and, specifically, in the $Z$ and $W$~boson mass measurements. This large luminosity can also be exploited to study the final state polarisation using $\tau$ leptons produced in the final state.

Concerning the topic of strong interactions, 
a number of people advocated in favour of a rich QCD programme during the discussion. For this kind of research, the ECRs were in favour of a hadron-hadron or hadron-lepton collider, in particular FCC-hh and/or FCC-eh. 
The latter will provide the complete resolution of parton distribution functions required for the subsequent hadron-hadron runs.
Nuclear structure, hot and dense QCD and heavy-ion physics can only be performed in hadron colliders, and ECRs were in favour of FCC.
The topic of collectivity in proton-proton and proton-nucleus collisions is regarded by some ECRs as the biggest discovery at the LHC besides the Higgs boson. 
However, some precision QCD measurements can also be performed with CLIC.

The diversity of experiments is also important to provide crucial tests of the Standard Model at various energy scales. 
Nonetheless, about 42\% of survey respondents do not believe that large long-term projects negatively impact the diversity of the field. 
As such, while diversity of experiments should be maintained, this should not negatively impact the development of future larger-scale projects.

On the phenomenology side, the ECRs were in agreement with the statement reported in the \BB~about the need for improving the event generators. It is of great importance to the ECRs that effort is made to have experimentalists and theorists interact more.
There is some level of consensus that at the moment it is hard to know where best to search for new physics in EW sector, and it is felt that joint workshops/conferences would help improve the situation in the future. 
It was discussed whether particle physics should move from benchmark model driven searches to final-state driven searches in expanding phase space. 
There was some support for increased use of Effective Field Theory (EFT) interpretations over benchmark models in physics analyses, though this was not universally supported.

\subsubsection{Comments on the Briefing Book}
The ECRs broadly held the view that direct, model-independent searches, as well as precision measurements and indirect searches are appealing approaches. 
There was mention of the $\kappa$ framework~\cite{kappaframework}; however, some of the ECRs thought that EFT interpretations were not given the space they deserve given that they are a quickly-developing topic in particle physics, and are expected to become more prominent in the years to follow. 
Further to this, the ECRs question the current level of focus on benchmark scenarios.
The \BB~summarises the physics, while avoiding drawing any conclusions, and does not show any studies where the expected results for the FCC-ee and the FCC-hh are given separately from each other, which is crucial to understand the physics reach of the individual FCC phases. 
Also, there is very little mention of projections from the HE-LHC and LE-FCC projects.  Finally, in the QCD chapter there is no mention of using polarised beams. The \BB~highlights lattice QCD as an important tool for precision tests of the Standard Model, a view also shared by the ECRs. 
It should be noted that dispersive techniques, which e.g.~currently give the most precise determination of the hadronic contributions to the anomalous magnetic moment, are left out from the \BB.

\subsubsection{Summary}
In summary, an $e^+e^-$ machine would give us the chance to perform a variety of precision measurements and searches in the electroweak sector, while allowing the total width of the Higgs boson to be measured. 
It would also benefit from a shorter timescale. 
However, lepton colliders seem to be slightly disfavoured by few ECRs in the discussions regarding strong interaction physics. 
A high-energy hadron-hadron collider offers a complementary physics programme, which would successfully execute most of the major goals in the fields of electroweak and strong interactions, including an improved measurement of the Higgs trilinear coupling, but with the exception of measuring the total width of the Higgs boson. 
However, it is clear that the time scale of the entire FCC programme is much longer than the other straw-man scenarios.  
While the ECRs strongly support collaboration and participation in international projects, the survey also showed that the ECRs consider a future European collider after the HL-LHC very important. 

\subsection{Beyond Standard Model, Dark Matter, and Dark Sector Physics}
Given there is no obvious sign of New Physics, our biggest challenge is to maintain excitement for BSM searches, both in the ECR community and in the public sector. Of note though, amongst the full set of delegates, Dark Matter is cited as being the most important and most interesting open question in HEP~(Figure~\ref{fig:survey_ecfa_q3}).

Beyond the Standard Model, Dark Matter, and Dark Sector searches have a vast landscape. It is noted that input to the strategy was provided by selecting benchmarks and comparing the performance of candidate experiments. This brought up the following questions amongst the ECRs:
\begin{itemize}
\item Is working with a few model benchmarks sufficient?
\item Would model scans be more effective in sampling the BSM landscape?
\item Should the particle physics community spend resources to find other more advanced and economical ways to provide a more robust input for BSM searches? 
\end{itemize}
Linked closely to the model strategy was the idea to bring various groups closer. Examples include communication and collaboration between both theorists and experimentalists as well as between different experiments (e.g. collider, direct detection, indirect detection). 

\subsubsection{Comments on the Briefing Book}
It was noted that the LUX experiment~\cite{Lux} as well as BSM searches with $\tau$ neutrinos and supersymmetry at SHiP~\cite{SHiPTP, SHiPPP} were missing in the \BB. 

Regarding different experiments, the majority of ECRs interested in this area were in favor of diversification of resources, or supporting a future collider that provides the opportunities for diverse experimental programmes, similar to how the LHC has been a part of a wider scientific programme encompassed by CERN. Concerns were raised over a single ambitious project with a long timescale (careers, motivation of younger generations, continuity of the field). The understanding from the \BB~was that current projections offer no obvious answer for the choice of the next-generation experiments. Complementary studies studies between collider, direct detection, and indirect detection experiments are important, especially for Dark Matter and Dark Sector physics.

\subsubsection{Need for Collaboration}
Particular focus was on the need for collaboration between theorists and experimentalists, and a strong desire was voiced for workshops with fewer talks, mainly dedicated to brainstorming and forming of collaborations (within or between experiments, TH-PH-EXP, etc.). The examples of the valuable outcomes of workshops such as Les Houches PhysTeV, LHC Hackathons, and workshops in direct detection experiments were discussed. The hope is to create awareness in funding agencies about benefits of such workshops or schools - particularly aimed at the PhD and postdoctoral level - as funding is often only given to people giving a talk, rather than participation in a workshop-style environment.

Experimentalists stressed the importance of encouraging new approaches to searches, streamlining search times, and more collaboration with theory colleagues for well-motivated models. Theorists, meanwhile, emphasised the importance of improving the presentation of results for reinterpretation and better communication with experimentalists.  

\subsubsection{Community Engagement}
Finally, the community was keen on outreach, education and communication. Noted was the success of outreach activities dedicated to a specific subject, e.g. Dark Matter Day~\cite{DarkMatterDay}. Additionally, the topic of making open data more accessible was viewed favourably, especially through finding creative uses such as work with high school teachers and students and ``citizen science'' projects, especially related to collider physics, such as CRAYFIS~\cite{Crayfis} and CREDO~\cite{Credo}. 

\subsubsection{Summary}
While the lack of evidence of New Physics is a challenge to the Beyond Standard Model, Dark Matter, and Dark Sector Physics community, the open questions stemming from these areas remain among the most interesting and important questions to HEP. A theme emerged of needing closer collaboration between not only the different experiments but also between experimentalists and theorists. Additionally, the ECRs in this group were highly motivated through involvement in community engagement and education. 

\subsection{Flavour, Neutrino and Cosmic Messenger Physics}


\subsubsection{Comments on the Briefing Book}\label{pbbflavnucosm}

The general consensus of the ECRs regarding the \BB~was that the goals of the physics topics discussed in chapters relevant to this section (i.e. Chapters 5-7) were well laid out.
Due to the conciseness of the \BB, it is understandable that the full breadth of the physics programme explored by some of the experiments was hard to capture. 
In this sub-section, a brief list of topics not covered in the \BB\, but relevant to the ECRs is presented.

In the heavy-flavour sector, the studies of the decays that undergo $b\rightarrow c \ell \nu$ and $b\rightarrow s \ell \ell$ transitions have turned out to be extremely interesting. These decays, along with Dark Matter searches, involve final states with missing energy. 
It would be appealing to have estimates on the reconstruction performances for such final states at the different future colliders/experiments.
The flavour-changing decays of $c$ and $b$ baryons have been explored and such systems provide a complementary ground to search for New Physics contributions. The \BB~mentions the measurement of electric and magnetic dipole moments of heavy baryon systems and $|V_{ub}/V_{cb}|$ extraction through $\Lambda_b$ decays. However, it ignores some potentially interesting topics that could be probed with baryon systems, including the potential discovery of CP violation, tentative flavour anomalies in the baryon sector, etc.
The importance of studies of charm decays is evident from the \BB. In light of this, it would be interesting to have estimates of sensitivities to rare charm decays and mixing parameters at the proposed future experiments. 
The measurements of the electric dipole moment of the $\tau$ lepton have been discussed in the \BB. However, there is no mention of the $\tau$ magnetic dipole moment, although there have been several proposals to measure this quantity at different future experiments with various techniques.

In the neutrino sector, the ECRs felt that the \BB~could have benefited from a bit more detail on the $\nu_\tau$ sector. The measurements of $\nu_\tau$ appearance and/or disappearance provide stringent tests of unitarity of the neutrino mixing matrix (PMNS). Although $\nu_\tau$ appearance is mentioned in the ``Cosmic Messenger" chapter in the section about the IceCube experiment, the prompt $\nu_\tau$ production at the hadron colliders could also be considered. To facilitate these studies, measurements of the $\nu_\tau$ cross-section from experiments such 
as DsTau, NA-65 and SHiP would be extremely helpful. These cross-section measurements can also lead to better understanding of the $\nu_\tau$ backgrounds in the next generation of long-baseline experiments and astrophysical $\nu_\tau$ detection in neutrino telescopes.  
Of interest in the neutrino sector are also solar neutrino experiments, coherent scattering experiments and searches for exotic BSM effects.
Topics such as Lorentz invariance violation, quantum decoherence, quantum gravity or large extra dimensions could be tested using long baseline experiments, while at low energies, measurements of magnetic moment or millicharge of neutrinos can be performed.

Regarding the Cosmic Messengers chapter, it was mentioned in the debate that the ``Synergies with HEP" section could emphasise new accelerator measurements of hadronic interactions that can help to improve the modelling of cosmic ray air showers. 
In this regard, the data from proton-oxygen and proton-nitrogen collisions would play a vital role in reducing systematic uncertainties in the simulation of such interactions and would lead to a better understanding of the backgrounds for the measurement of high-energy cosmic leptons. 

\subsubsection{Future of the Field}\label{futureflavnucosm}

Generally speaking, the heavy flavour sector can benefit from all of the five future scenarios considered in the \BB. However, the light, neutrino and cosmic messenger
sectors require dedicated experiments. 

In the heavy sector, a comprehensive study of the sensitivities to the observables related to $c$ and $b$ hadron decays for all the proposed scenarios is not available.
Studies on flavour tagging performance for different hadrons and on the reconstruction performance of decays with missing energy would be of particular interest. Considering the FCC scenarios, the FCC-ee provides a clean environment to study heavy flavour physics.
This, however, is not as clean as the environment in colliders operating at the $\Upsilon(4S)$ threshold, but it does facilitate production of different $b$-hadron species. The FCC-ee will also improve our knowledge of $\tau$ decays and $B$ decays involving $\tau$ leptons in the final states.
In the global context, according to the available studies, with regard to the heavy-flavour sector, similar performances are expected between FCC-ee and CEPC experiments.
The FCC-hh option would lead to higher production rates for $c$ and $b$ hadrons. However, the extreme collimation of the decay products at these energies, together with the very high levels of pile-up, make it rather unclear whether heavy-flavour physics can be actually performed properly in this scenario. 
The \BB~does not discuss in detail the heavy-flavour physics at CLIC and ILC, although both projects 
are considering having a dedicated part of the programme for collisions at the $Z$-pole, and both projects
have studied in detail the capabilities of the flavour-tagging and quark-charge measurements. This fact made discussions during the debate difficult and we hope that this is better covered in future reports.
With the sensitivity studies shown in the \BB~regarding flavour physics with Higgs, top quark and $Z$-boson, ECRs recommended starting with either an $e^+e^-$ collider (such as FCC-ee or CLIC) or electron-hadron collider (such as LHeC) and later move towards a hadron-hadron collider. In the light sector, it is evident from the \BB~that dedicated experiments are needed to the achieve best sensitivities to observables such as electric dipole moments, lepton flavour violating (LFV) observables, etc. 
All the current and future charged LFV experiments such as MEG II \cite{MEGDetector}, Mu3e \cite{mu3e}, Mu2e \cite{mu2e} and COMET \cite{COMET} are based outside of CERN. These experiments are limited by the muon beam rate and the detector technology. An improvement in these two fields may partially come as a by-product of R\&D on large-scale experiments. 
For kaon physics, as highlighted in the \BB, dedicated experiments will be required, and future hadron colliders would be highly beneficial in this regard. 

In the neutrino sectors, two long-baseline experiments are currently planned, one in the USA and one in Japan. European researchers are strongly involved in both experiments. The expertise gained through these projects can be used to set up a next-to-next-generation long baseline experiment in Europe. Nonetheless, plenty of perspectives exist in Europe already for neutrinoless double-beta decay and neutrino mass measurements.

In the cosmic messenger sector, real-time observations between connected observatories, for example neutrino, gravitational wave and gamma ray telescopes, will be crucial in the future to inform each other regarding the transient events.

\subsubsection{Summary}\label{summaryflavnucosm}

From the meeting, there was no concrete agreement on a final recommendation for the preferred collider scenario. 
A majority of the ECRs, however, did feel that an option of FCC-ee in the beginning and then FCC-hh would be beneficial for heavy-flavour physics. 
Other social aspects were also discussed in the meeting, with the following deemed particularly important by the ECRs: the diversity of the research field given that there is no clear path towards New Physics; cross-functional collaboration between experiments to share ideas, techniques, resources, etc.; and platforms for strong collaborations between experimentalists and theorists.
\subsection{Accelerator and Detector R\&D}

Large consensus was found among ECRs on the importance of Europe constructing a next-generation collider after the HL-LHC.
Moreover, about 88\% of ECRs believe that the next collider should be an $e^+e^-$ machine although there is no clear preference if this machine should be linear or circular.
Therefore, a strong and diverse R\&D programme on both accelerators and detectors must be a high priority for the future.
According to the survey, about 40\% of the invited young researchers to the ECR meeting are involved in R\&D activities: 35\% for detector and 5\% for accelerator R\&D.

\subsubsection{Comments on the Briefing Book}
\label{det_bb}

An introduction about different technologies for both future R\&D accelerators and detectors is given in the \BB, but a detailed description of the state of the art is missing.
Moreover, the distinction between first-generation and second-generation future colliders and the level of maturity of each accelerator proposal are not clearly described in the document.

In particular, concerns have been raised on the accelerator side as to whether the key numbers stated in the \BB~allow for a fair comparison of the various projects.
For example in terms of precision physics, different simulation tools with different levels of precision and maturity (fast simulation vs. full simulation) were used to evaluate the physics projections.
Also, no uncertainties were included in the estimation of financial aspects and timescales for the R\&D, the construction and operation phases of different future projects that would reflect the different level of maturity of the projects themselves.
Furthermore, the ECRs criticise a lack of risk assessment related to a possible delay of the projects.
Neither was the environmental impact of different future projects discussed, whereas 73\% of the ECRs agree that this topic should be thoroughly evaluated when making decisions on future projects.

On the detector side, there was no mention of which technology is suitable for which future project.
The readiness of each technology to cope with the challenges set by the different environments and analysis requirements remains unclear.

There is also no mention of the extent to which the decision process of other countries will influence the European particle physics landscape.
In particular for the 2020 ESU, it would be important to include in the final recommendation the consideration of the different scenarios that could appear if the ILC or CepC are approved by the Japanese and Chinese governments.
We regret that this was not done for the \BB~and
we argue that the ESU should state clearly the different options of CERN participation, possible support and synergies with other international projects. 
In fact, we believe that CERN, as a world-leading laboratory in particle physics, should play a major role not only in CERN-based future colliders but also in other international projects. 
The particle physics field -  and more specifically the R\&D it necessitates - has become an international activity.
We therefore argue strongly for a continued  strengthening of the collaboration between European and non-European particle physics communities.

\subsubsection{Diversity}
\label{det_diversity}

There is a number of smaller experiments exploiting the LHC proton beam to search for BSM and more exotic physics, complementary to the main large experiments.
Interest in running such experiments is currently growing, and we stress the importance of such diversity in part due to the lack of clear evidence for the existence of BSM physics.
It is clear that these diversification efforts are required now and cannot come at a later stage.
Furthermore, taking advantage of diversity should be part of the conceptual design of future colliders, laying out what number and kind of experiments could run at a future machine in parallel to the main ones.
This is crucial for the first generation of future colliders, but should also be part of long-term concepts as a plasma-based collider or a muon collider.
We regret that this topic has not been covered sufficiently in the \BB. 
It is not excluded that large long-term projects can be useful to attract expertise and funding at CERN and therefore have an overall positive impact on the diverse programme, but we strongly argue that the ESU must include these studies for each proposed future accelerator.

\subsubsection{Need for Equal Recognition} 
\label{det_social}

Among young researchers, there is a growing concern about the difference of career prospects between individuals pursuing mainly R\&D tasks and mainly analysis tasks.
This concern is clearly expressed by the ECRs  where only about 10\% strongly agree with the statement \textit{``I think Detector R\&D and Computing work is well-recognised in my experiment/group"} and it was also expressed by ECRs involved in computing, object reconstruction and performance measurements.
The issue of unequal recognition in comparison to particle physics analysis is
especially concerning for the success of the future of experiments, where a large degree of specialisation is more and more needed.
We encourage the creation of diversity in tasks within one experiment and to establish equal recognition not only within the group and experiment but also in universities, laboratories and, more generally, in academia.
We support the introduction of specialised career paths/professorships at universities and labs and awards for all types of specialisation, with emphasis on the technical aspects such as detector/accelerator R\&D or computing.  
We also want to underline the importance of training and of having more people that can bridge the gap between technical tasks (computing/software and R\&D) and physics analysis mentioned in the \BB.

The possibility of introducing separate career tracks not only for physics analysis and R\&D (or computing) but also for professional/expert tasks and more management-focused tasks was discussed. 
We believe that this could be done in sufficiently large groups or institutes and should be addressed in the ESU. 
This separation could improve research efficiency, while making such tracks more attractive.

\subsubsection{Summary}
In summary, it is agreed that the ESU should emphasise, much stronger than the \BB~and the 2013 ESU do, that any future small or international project must address environmental issues and career prospects much more seriously and thoroughly.
We also believe that explicit emphasis should be put on work-life balance improving the family-friendliness and attractiveness of the field. 

\subsection{Computing and Software}

\subsubsection{Comments on the Briefing Book}

In the \BB, software and computing in HEP is covered in the
instrumentation chapter along with particle detectors. Particular emphasis on the computing side is put on the immediate challenges posed by the
HL-LHC with its unprecedented data output. The budget for computing is expected to
stay constant, so a vast R\&D programme in an environment that fosters innovation
is necessary to meet future needs.

Another focus of the \BB~is the link between instrumentation and computing. 
To maximise the physics potential and to make best use of the monetary and human resources in HEP, the impact on computing should be increasingly considered in detector design. 
This requires individuals which can bridge both communities.

The \BB~also acknowledges the merit of software research and
development, as it might allow for ``tool-driven revolutions'', and emphasises the
importance of networks and organisational structures that extend beyond the HEP
community to make use of synergies with industry and other fields of physics or research in general.

We agree with the key points mentioned in the \BB~and in particular with the observation that software and computing activities must be recognised not only as a means to do physics analyses, but as research that requires a high level of skill. 
Therefore, this section mainly touches on other subjects which were less prominent in the \BB.

\subsubsection{Software for Physics Analysis}

The \BB~talks about ``computing in HEP'' as mainly the central
production processes (e.g. simulation, reconstruction, data management), but we feel that software and computing also goes closely with physics analysis.
Innovation in this area should strive for minimisation of the time that it takes
to get to a physics result and for the redirection of this economised time and
person-power to other places where innovation and development is truly needed.
One way to achieve this is the centralisation of selected analysis tasks within the collaborations to reduce duplicate efforts in different analysis groups.

Rethinking how physics analyses are done can also mean building software
tools that are easy to understand and can be used for multiple purposes.
However, care should be taken to avoid ``reinventing the wheel'' and to make use
of standard software packages where available. This is very important to ensure
the acquired software skills are transferable to other areas outside HEP. For
example, one should leverage on software trends like continuous integration to
automate analyses.

Encouraging more streamlined analysis workflows does not come without risks.
Centralising analysis workflows often allows for reduced data formats to save storage space and analysis execution time. 
Precautions must be taken that this does not complicate the execution of more unconventional analyses, i.e. machine learning (ML) approaches which aim to directly analyse low-level detector data. 
It is also important that any change in the analysis workflows does not increase the separation between the
physicists working on physics objects and those working on physics analysis.

On a more general note, many ECRs feel it is important that experts contribute to HEP software for sample production and analysis to ensure most
efficient use of resources and minimise time to produce physics results. Physicists should be
encouraged to become experts with training or by self-learning. 
Another possible option to ensure software quality is collaborating with computer scientists outside particle physics.

\subsubsection{Analysis Preservation and Open Data}

During the ECR plenary discussion, quite some time was spent on the subject of
analysis preservation and open data. Analysis preservation means that published
analyses are documented and archived in such a way that it is possible for future
generations to reproduce paper results. The importance of good documentation for
analysis software and scripts should be self-evident, but it becomes
increasingly important as new tools are on the rise which are not necessarily
familiar to everyone in the community. Collaborations could make well documented analysis
software a requirement for their papers, and clear documentation standards
inspired by industry could be adopted. In particular, the usage of complex
analysis algorithms like machine learning poses new challenges to
reproducibility.

Although not exactly related to software and computing, we want to stress the importance of providing supplementary material along with papers. Experimental results should be available in a form that can be
easily reinterpreted by theorists. This means in practice that not only plots and data directly presented in papers should be easily accessible, but also additional results, as well as trained machine learning models, that might be of interest to the expert but are not included in any publication for reasons of brevity. We encourage the use of HEPData~\cite{HEPData} or
toolkits like Rivet~\cite{Rivet} to achieve this goal. 
We encourage to initiate more projects like the CERN Analysis Preservation Portal~\cite{analysisPreservation}.

Open data means that the data (and the software to analyse it) should be freely
available to everyone to use and publish new results as they wish without
restrictions. In the survey, 92.4\% of the participants indicated they are generally in favor
of open data, with approximately 60\% of the participants thinking the primary
purposes should be education and scientific goals. The third option, outreach,
followed with roughly 40\%.
Of those in favor of open data, 76.3 \% agreed that there should be
an embargo period for the data before it is released. 
There are several concerns regarding publishing data immediately. Dubious analysis results by non-experts could have a negative impact on the reputation and reliability of the experiments that collect the data. While it is agreed that there is value in interacting with other fields of science and technology there are doubts about how useful publications from non-physicists would be from a scientific perspective. Finally, publishing data before it is analysed by the corresponding collaborations is incompatible with the way particle physics experiment currently operate. It would drastically change the role of analysers within the collaborations, as scientists would be more inclined to write single-author papers. The recognition and incentive for the physicists building and running experiments would be undermined. The current experiments are based on collaborative work, where developments benefit the whole collaboration and in return every member is an author to every publication. There is no interest at the moment among the ECRs to fundamentally change the publication procedures.

\subsubsection{Innovation in High Energy Physics Computing}

We support software R\&D projects and organisations like DIANA-HEP~\cite{Diana}, IRIS-HEP~\cite{IRIS} or
the HEP Software Foundation~\cite{HSF} and the HEP community already greatly benefits from their work.
It is important to encourage knowledge sharing and collaboration not only with
industry but also with other fields of physics or science in general and to detect possible synergies.
More efforts should be undertaken to leverage already existing software solutions and learn from other approaches to similar problems. 
There also needs to be wider awareness in the HEP community of the methods used by other fields in physics. 
Astrophysics for example often faces similar challenges related to the data rate, but has different dynamics in how they are tackled due to the shorter timescales of experiments (relative to collider experiments) and the different organisation structure.

\subsubsection{Work Recognition and Career Prospects}

Computing and software activities should not be seen only as a means to perform physics analysis but as a proper research area.
The visibility and recognition of the individuals contributing to software can be increased by awards inside the collaborations and by encouraging more publications on software and computing work.
The rise of heterogeneous computing requires adequate training of researchers at all career stages, as making use of emerging technologies asks for either highly-trained physicists or computer scientists (possibly even from outside HEP).
Job descriptions should be more explicit in the required software skills to increase awareness of in-demand expertise.
More specific career opportunities for computing-oriented physicists in universities and laboratories are necessary.



\subsubsection{Impact of Software and Computing on the Environment}

The use of new technologies in HEP computing (e.g. GPUs and FPGAs) should not
only be encouraged to address immediate challenges posed by the HL-LHC, but also
to reduce the ecological footprint of computing in HEP. Although this should be
put into perspective (considering that the power consumption of HEP experiments
is dominated by particle accelerators), it should not be forgotten that the
development of more efficient computing techniques is also a concern in other
areas of human activity. The transfer effects of new developments
are especially strong in the software and computing sector.

Other opportunities to have a positive impact on the environment are related to
existing and future computing facilities. For example, it is important to build
relationships to other local computing communities to consider the sharing of
resources, i.e. computing hardware or cooling capacity.

In an effort towards reducing the amount of travel, software for remote
meetings should be improved and specialised for large-community meetings. The
product should be capable of reproducing as many interactions that traditionally
require face-to-face contact as possible, like drafting ideas together on a
whiteboard or allow for social interactions. This goes with the caveat that
meeting in person can not be substituted completely, as social connection is very
important to build mutual trust and networks in the community. Improving the
remote experience via software might have a big impact on the researchers
propensity to travel, although it could already go a long way to make stronger
efforts towards a more enjoyable remote experience in general. Currently, this
is usually treated as just a minor matter in the organisation of larger-scale
events.

\section{Final Remarks}

Having been mandated to collect the views of early-career researchers on the European Strategy for Particle Physics update, this document summarises these views, drawing from the plenary discussion held at CERN in November 2019.
It includes the outcome of discussions in preparatory meetings, on the day itself and the results of a survey circulated after the meeting. 

The ECRs were very pleased to be engaged in a process that will influence the HEP landscape for the duration of our careers and beyond, and would like to thank ECFA for this opportunity. 
However, the point in time at which the mandate was given was very close to the end of the entire ESU process. 
We therefore strongly and unanimously express that ECRs should be included from the start of the process in future strategy updates.
Nevertheless, the discussions were held in a very good spirit and were enjoyable and constructive. 
It was unanimously agreed that such discussions should become a more regular occurrence and we recommend that a permanent ECR council should be established within ECFA.

The Executive Summary already gives an overview of the main outcomes of the discussion. 
Therefore, we close by re-iterating the areas where we feel that significantly stronger emphasis is needed: 
keeping the field attractive to the best minds; 
putting at higher priority issues of the environment and sustainability; 
and increasing the focus on sociological and human factors.

\clearpage
\appendix
\counterwithin{figure}{section}
\section{Detailed Survey Results}
\label{appendix:socialsurvey}

\subsection{General Information}
\begin{figure}[h!]
    \begin{center}
    \includegraphics[width=0.48\textwidth]{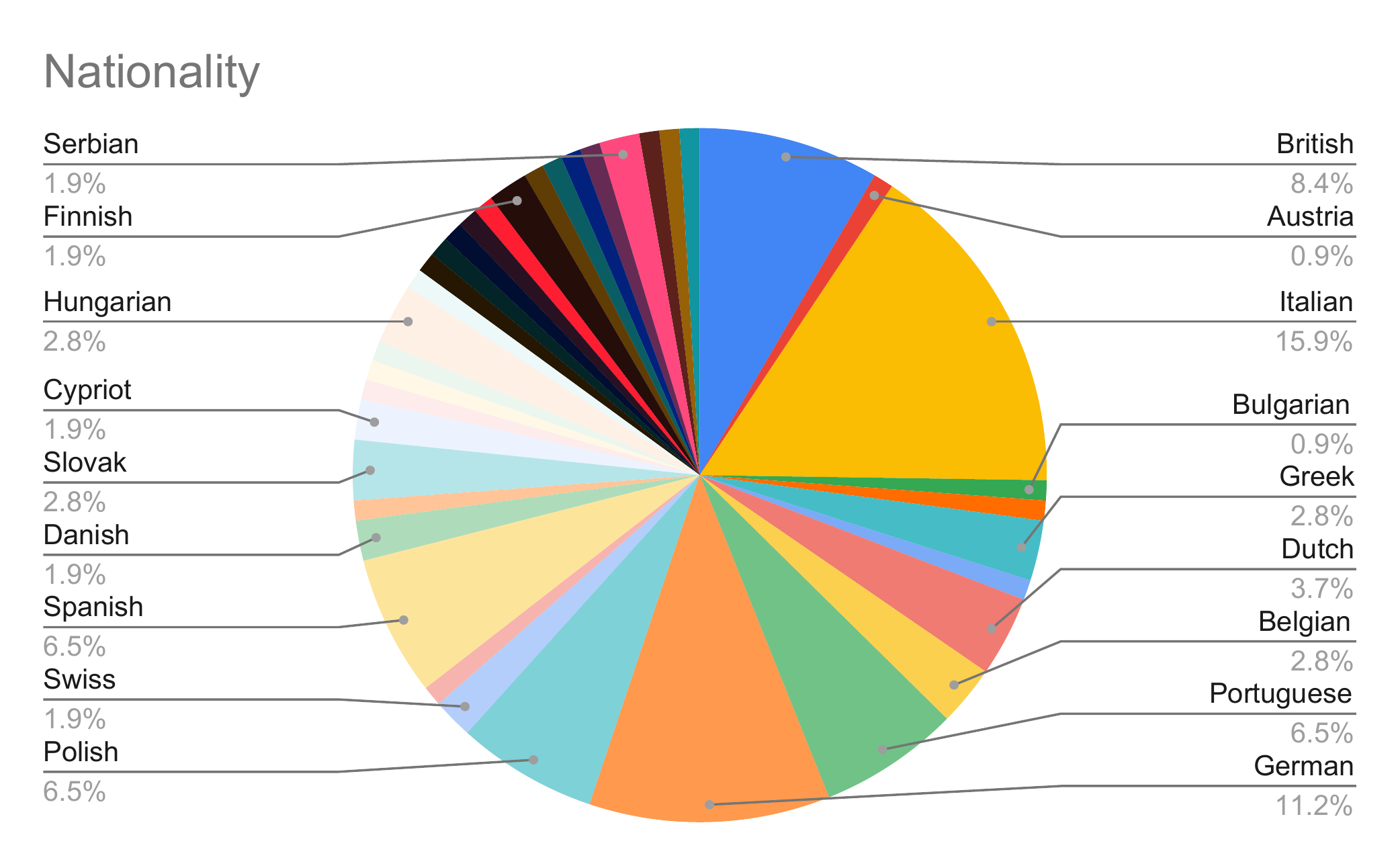}
    \includegraphics[width=0.48\textwidth]{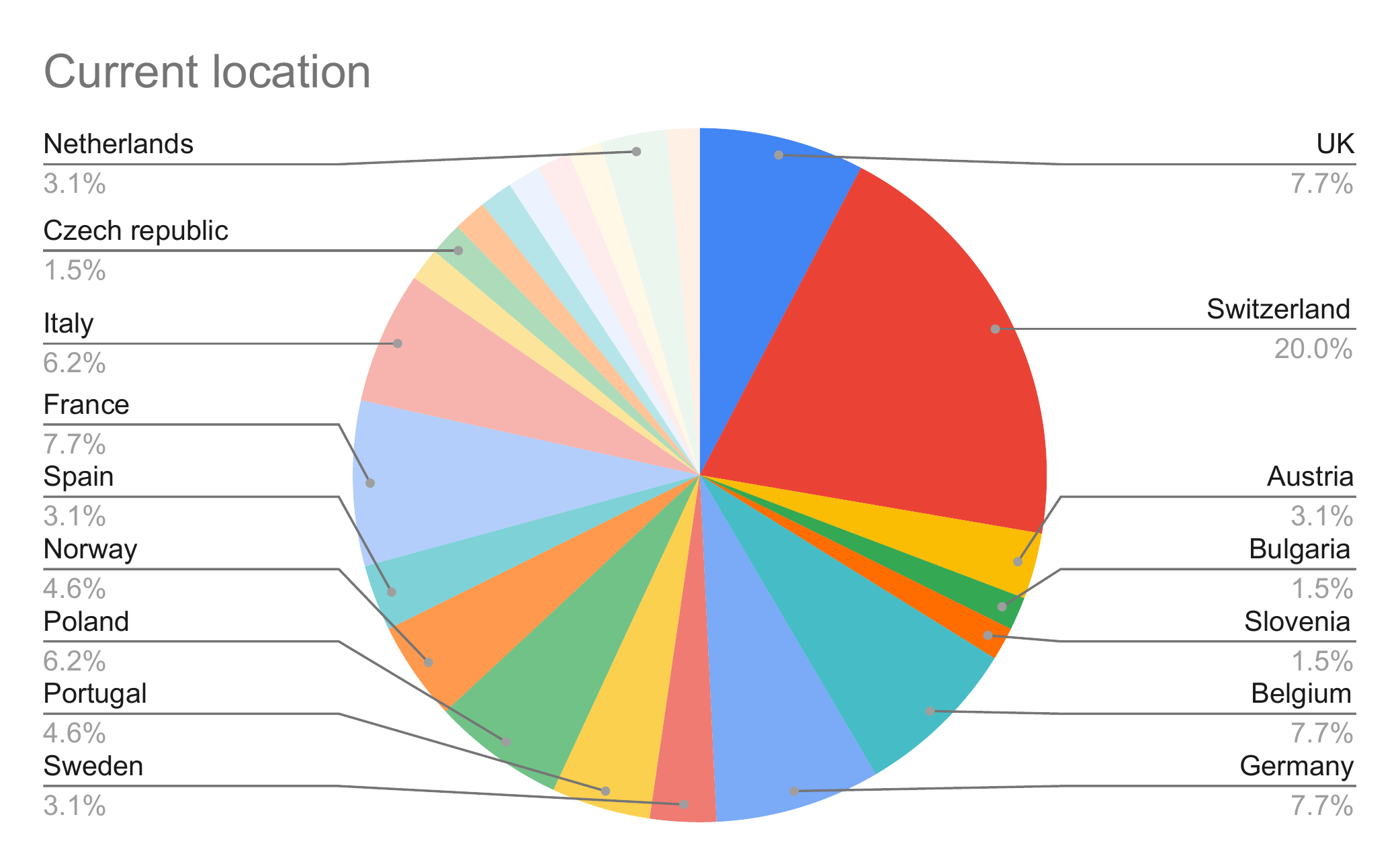}
    \end{center}
    \caption{Results of question 1 and question 2 of the general information section of the survey, which ask about the nationality and current location of ECRs.}
    \label{fig:survey_gen_q12}
\end{figure}

\begin{figure}[h!]
    \begin{center}
    \includegraphics[width=0.9\textwidth]{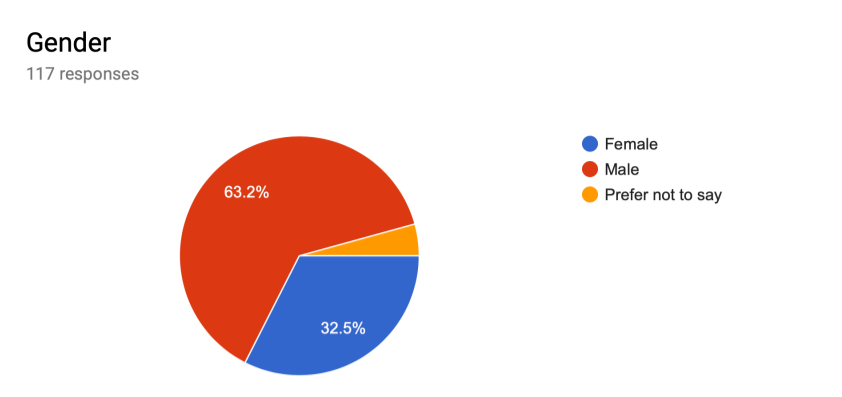}
    \end{center}
    \caption{Results of question 3 of the general information section of the ECR survey.}
    \label{fig:survey_gen_q3}
\end{figure}

\begin{figure}[h!]
    \begin{center}
    \includegraphics[width=0.6\textwidth]{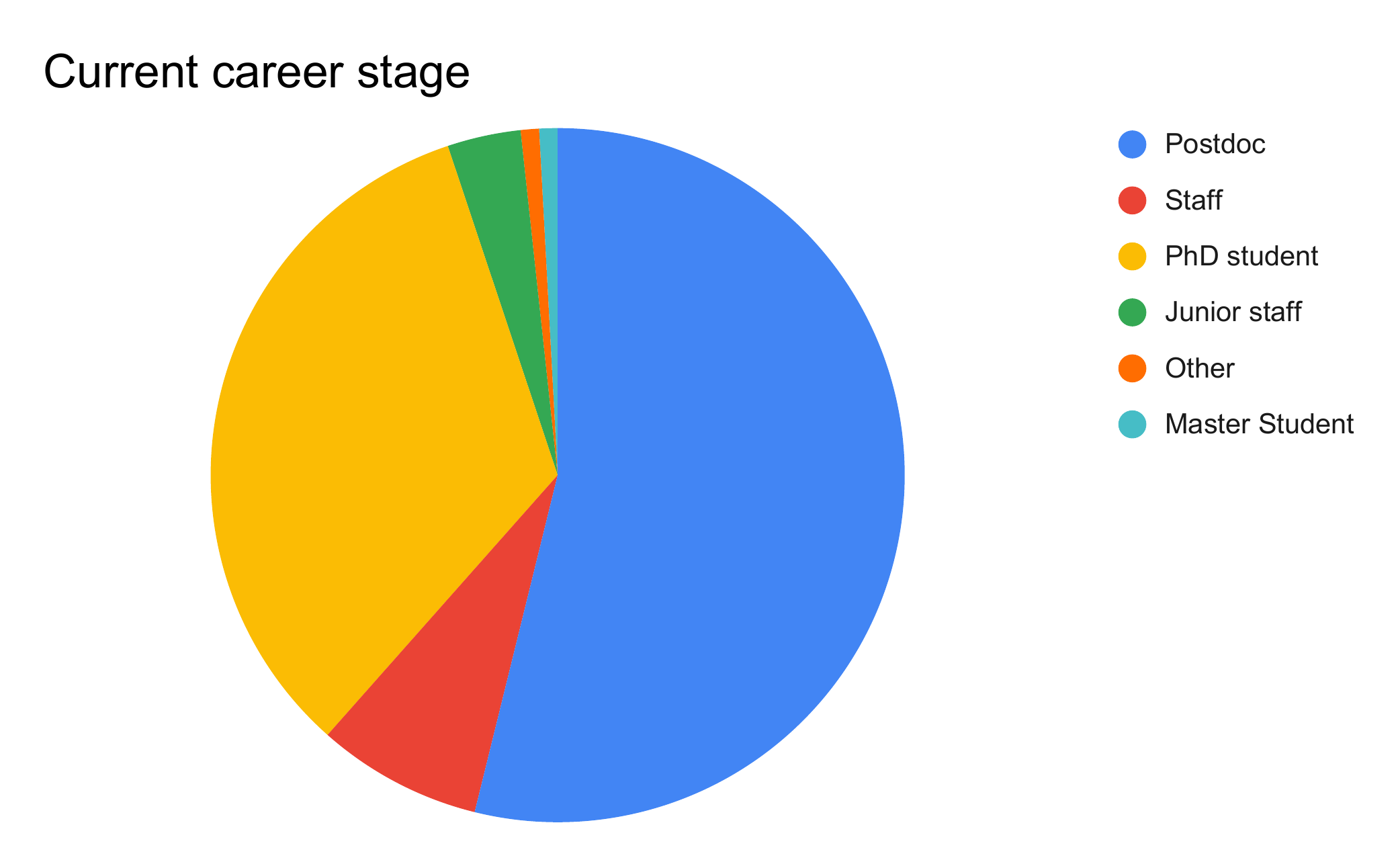}
\end{center}
    \caption{Results of question 4 of the general information section of the ECR survey.}
    \label{fig:survey_gen_q4}
\end{figure}

\begin{figure}[h!]
    \begin{center}
    \includegraphics[width=0.9\textwidth]{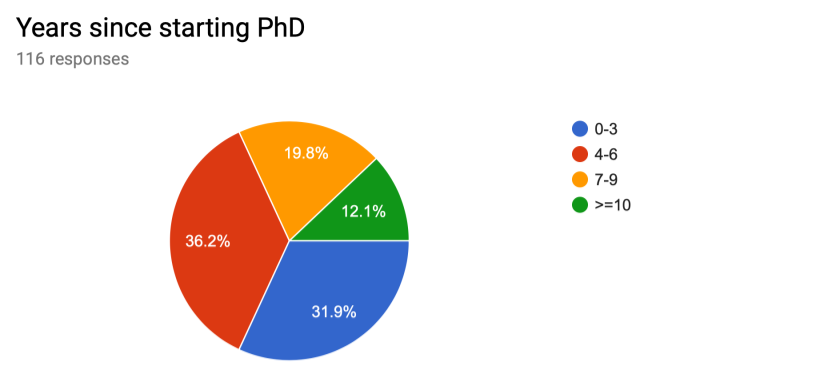}
    \end{center}
    \caption{Results of question 5 of the general information section of the ECR survey.}
    \label{fig:survey_gen_q5}
\end{figure}

\begin{figure}[h!]
    \begin{center}
    \includegraphics[width=0.9\textwidth]{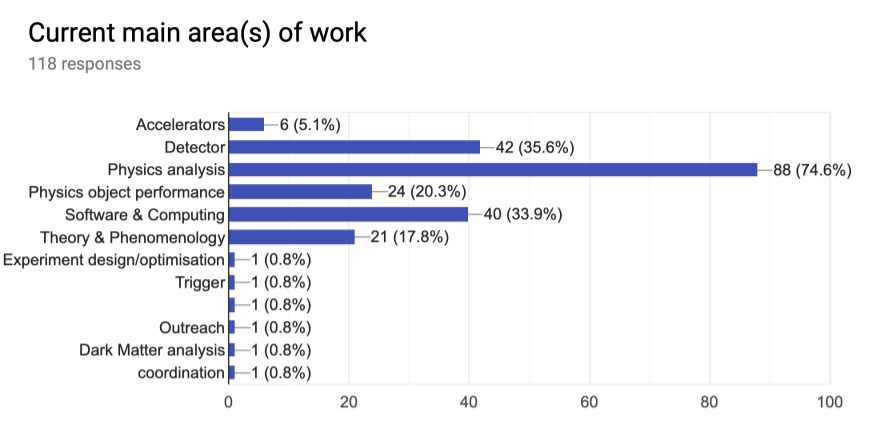}
    \end{center}
    \caption{Results of question 6 of the general information section of the ECR survey.}
    \label{fig:survey_gen_q6}
\end{figure}

\begin{figure}[h!]
    \begin{center}
    \includegraphics[width=0.9\textwidth]{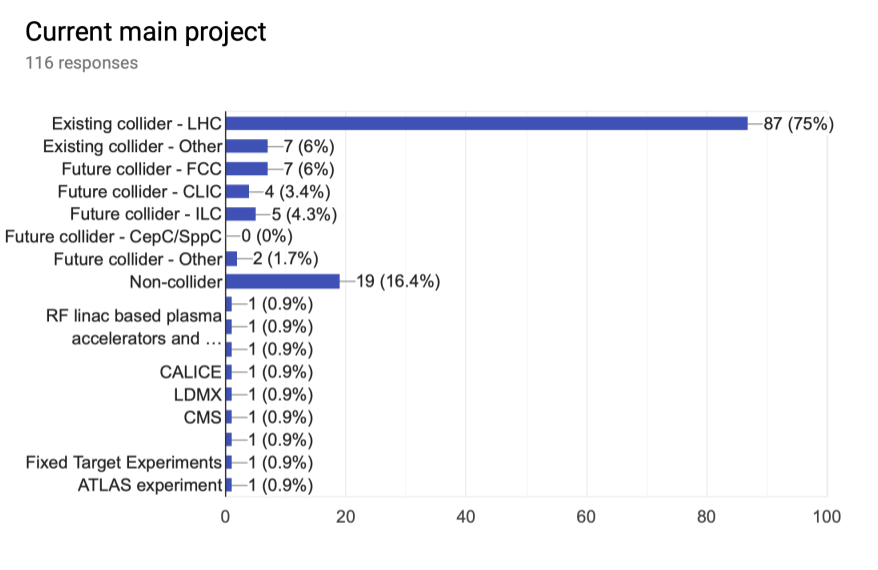}
    \end{center}
    \caption{Results of question 7 of the general information section of the ECR survey.}
    \label{fig:survey_gen_q7}
\end{figure}

\begin{figure}[h!]
    \begin{center}
    \includegraphics[width=0.9\textwidth]{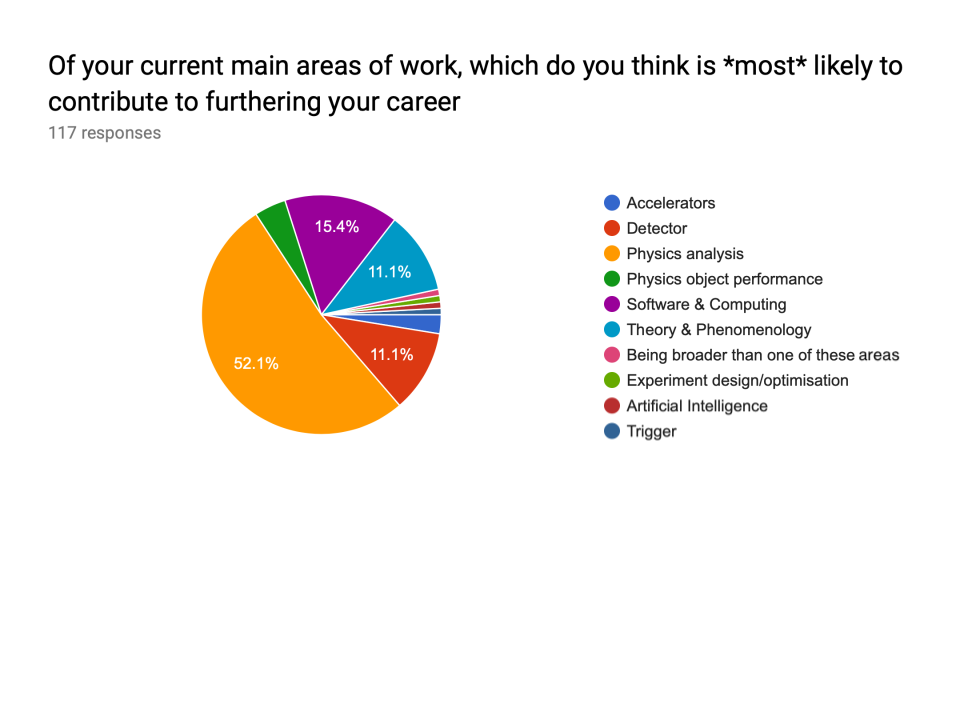}
    \end{center}
    \caption{Results of question 8 of the general information section of the ECR survey.}
    \label{fig:survey_gen_q8}
\end{figure}

\begin{figure}[h!]
    \begin{center}
    \includegraphics[width=0.9\textwidth]{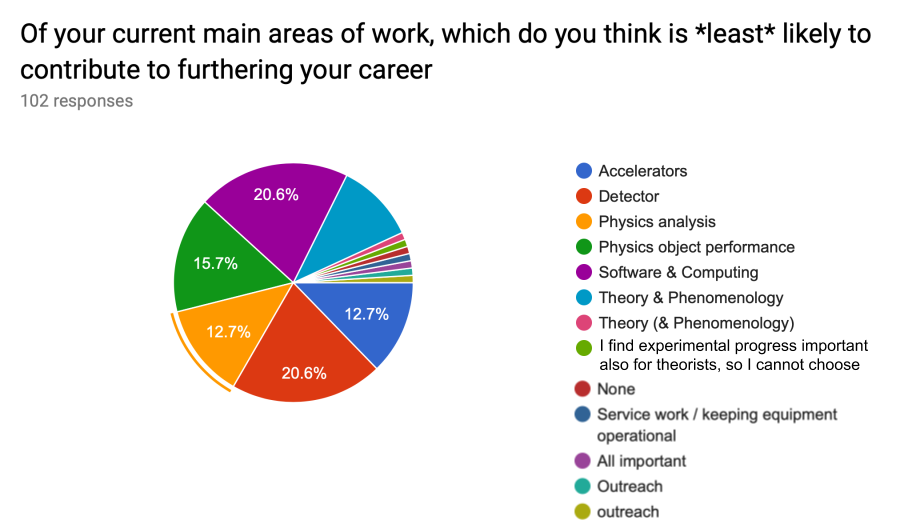}
    \end{center}
    \caption{Results of question 9 of the general information section of the ECR survey.}
    \label{fig:survey_gen_q9}
\end{figure}

\clearpage

\subsection{European Strategy Update}

\begin{figure}[h!]
    \begin{center}
    \includegraphics[width=0.9\textwidth]{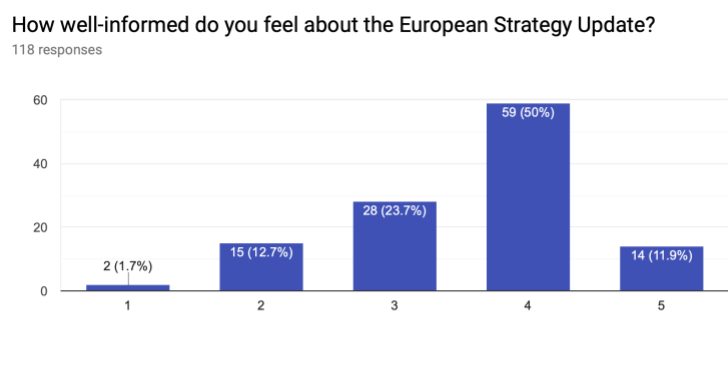}
    \end{center}
    \caption{Results of question 1 for the European Strategy Update section of the ECR survey, where (1) indicates strongly disagree, (2) disagree, (3) neutral, (4) agree and (5) strongly agree.}
    \label{fig:survey_ecfa_q1}
\end{figure}

\begin{figure}[h!]
    \begin{center}
    \includegraphics[width=0.9\textwidth]{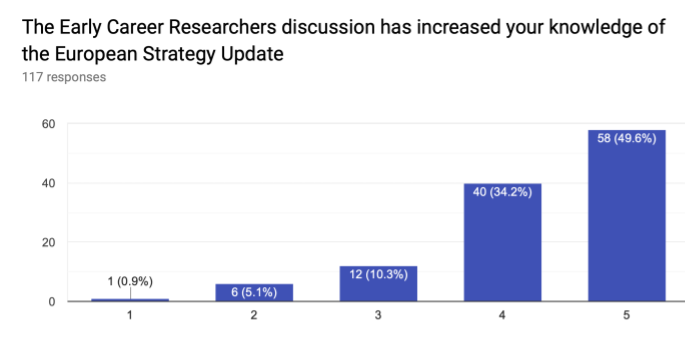}
    \end{center}
    \caption{Results of question 2 for the European Strategy Update section of the ECR survey, where (1) indicates strongly disagree, (2) disagree, (3) neutral, (4) agree and (5) strongly agree.}
    \label{fig:survey_ecfa_q2}
\end{figure}

\begin{figure}[h!]
    \begin{center}
    \includegraphics[width=0.9\textwidth]{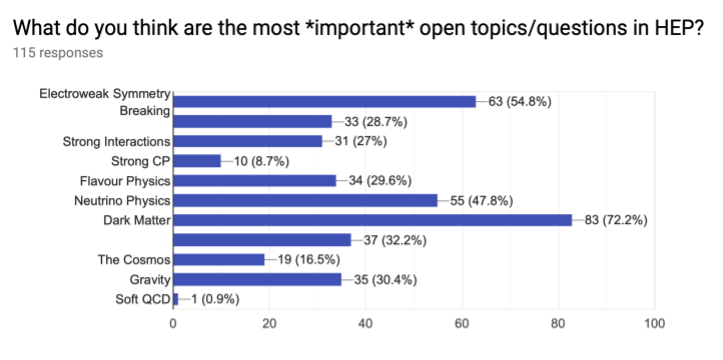}
    \end{center}
    \caption{Results of question 3 for the European Strategy Update section of the ECR survey.}
    \label{fig:survey_ecfa_q3}
\end{figure}

\begin{figure}[h!]
    \begin{center}
    \includegraphics[width=0.9\textwidth]{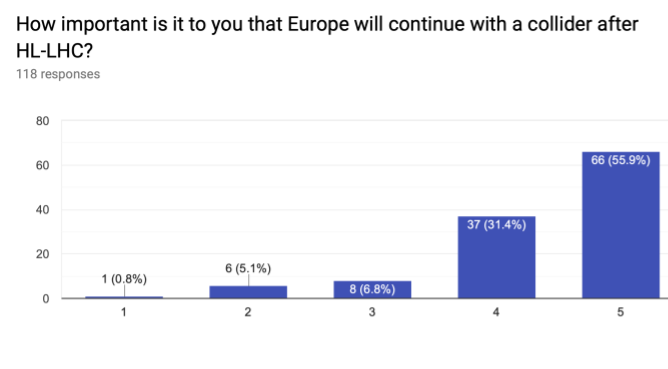}
    \end{center}
    \caption{Results of question 4 for the European Strategy Update section of the ECR survey, where (1) indicates strongly disagree, (2) disagree, (3) neutral, (4) agree and (5) strongly agree.}
    \label{fig:survey_ecfa_q4}
\end{figure}

\begin{figure}[h!]
    \begin{center}
    \includegraphics[width=0.9\textwidth]{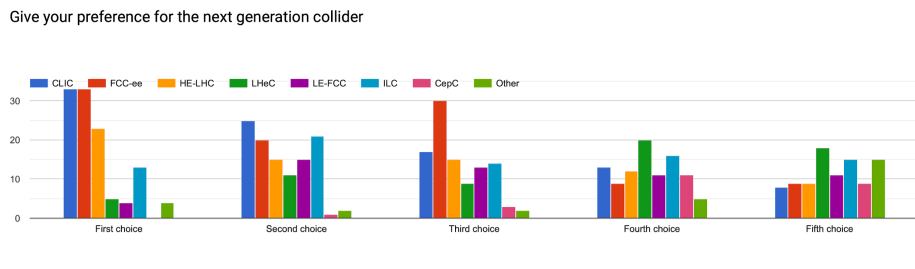}
    \end{center}
    \caption{Results of question 5 for the European Strategy Update section of the ECR survey.}
    \label{fig:survey_ecfa_q5}
\end{figure}

\begin{figure}[h!]
    \begin{center}
    \includegraphics[width=0.9\textwidth]{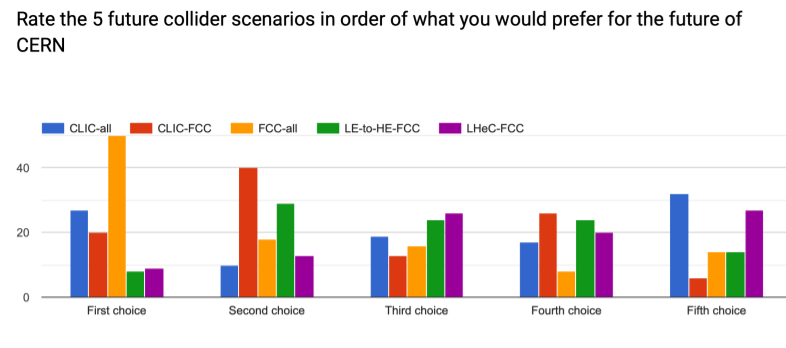}
    \end{center}
    \caption{Results of question 6 for the European Strategy Update section of the ECR survey.}
    \label{fig:survey_ecfa_q6}
\end{figure}

\begin{figure}[h!]
    \begin{center}
    \includegraphics[width=0.9\textwidth]{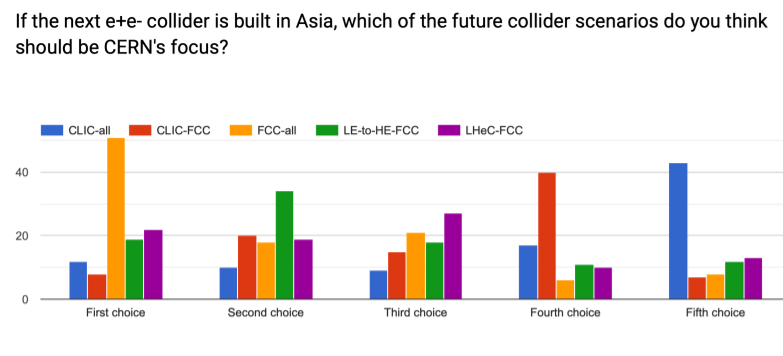}
    \end{center}
    \caption{Results of question 7 for the European Strategy Update section of the ECR survey.}
    \label{fig:survey_ecfa_q7}
\end{figure}

\begin{figure}[h!]
    \begin{center}
    \includegraphics[width=0.9\textwidth]{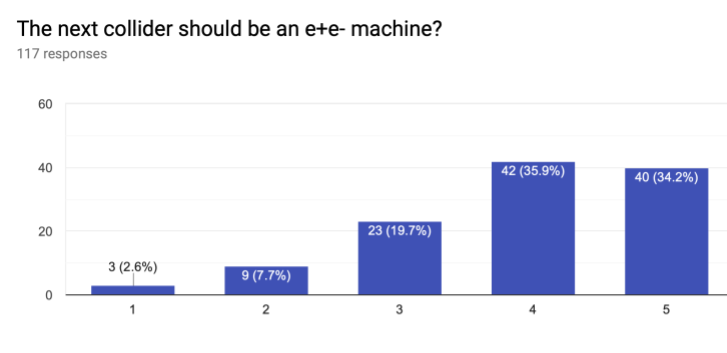}
    \end{center}
    \caption{Results of question 8 for the European Strategy Update section of the ECR survey, where (1) indicates strongly disagree, (2) disagree, (3) neutral, (4) agree and (5) strongly agree.}
    \label{fig:survey_ecfa_q8}
\end{figure}

\begin{figure}[h!]
    \begin{center}
    \includegraphics[width=0.9\textwidth]{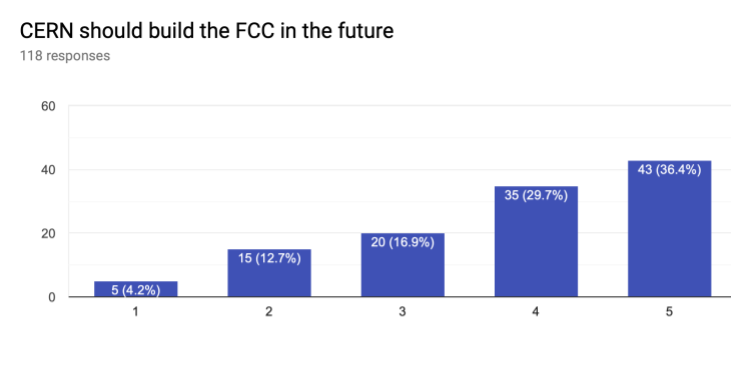}
    \end{center}
    \caption{Results of question 9 for the European Strategy Update section of the ECR survey, where (1) indicates strongly disagree, (2) disagree, (3) neutral, (4) agree and (5) strongly agree.}
    \label{fig:survey_ecfa_q9}
\end{figure}

\begin{figure}[h!]
    \begin{center}
    \includegraphics[width=0.9\textwidth]{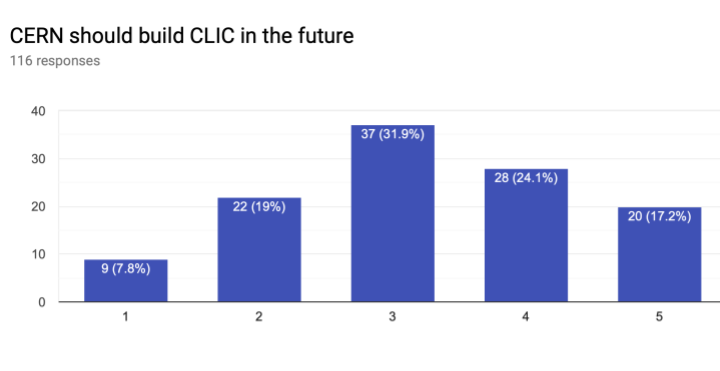}
    \end{center}
    \caption{Results of question 10 for the European Strategy Update section of the ECR survey, where (1) indicates strongly disagree, (2) disagree, (3) neutral, (4) agree and (5) strongly agree.}
    \label{fig:survey_ecfa_q10}
\end{figure}

\begin{figure}[h!]
    \begin{center}
    \includegraphics[width=0.9\textwidth]{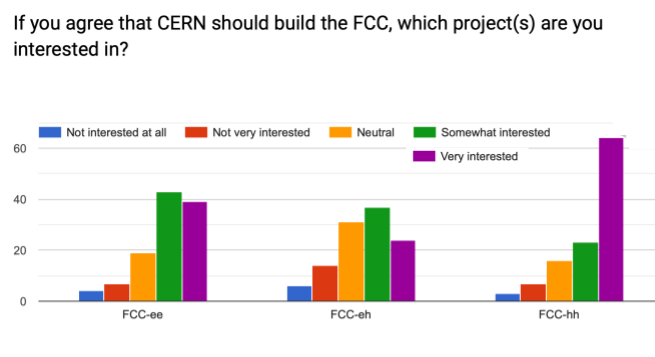}
    \end{center}
    \caption{Results of question 11 for the European Strategy Update section of the ECR survey.}
    \label{fig:survey_ecfa_q11}
\end{figure}

\begin{figure}[h!]
    \begin{center}
    \includegraphics[width=0.9\textwidth]{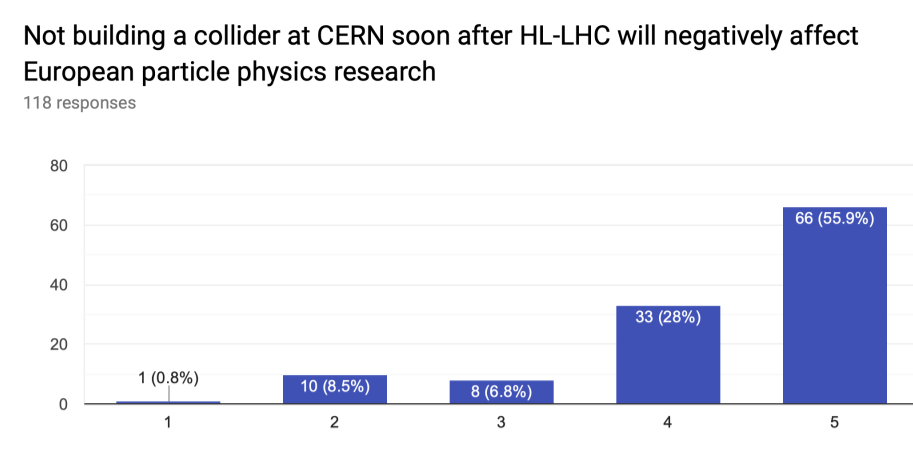}
    \end{center}
    \caption{Results of question 12 for the European Strategy Update section of the ECR survey, where (1) indicates strongly disagree, (2) disagree, (3) neutral, (4) agree and (5) strongly agree.}
    \label{fig:survey_ecfa_q12}
\end{figure}

\begin{figure}[h!]
    \begin{center}
    \includegraphics[width=0.9\textwidth]{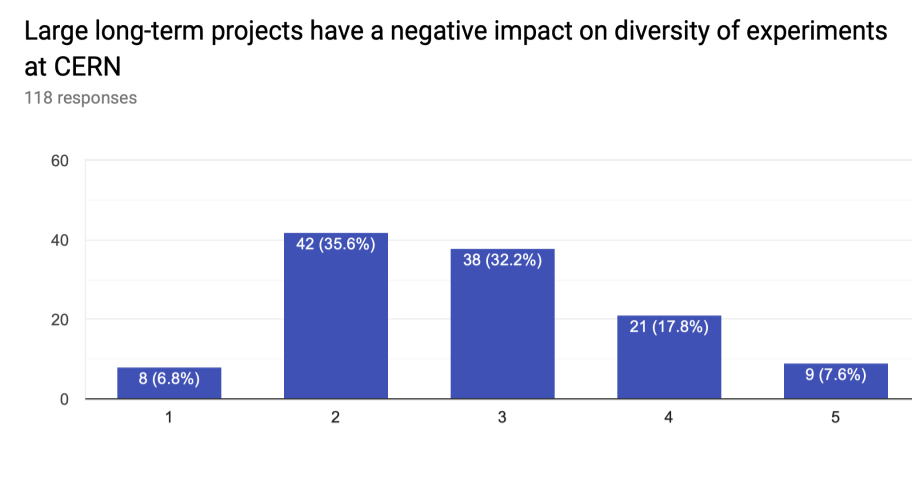}
    \end{center}
    \caption{Results of question 13 for the European Strategy Update section of the ECR survey, where (1) indicates strongly disagree, (2) disagree, (3) neutral, (4) agree and (5) strongly agree.}
    \label{fig:survey_ecfa_q13}
\end{figure}

\begin{figure}[h!]
    \begin{center}
    \includegraphics[width=0.9\textwidth]{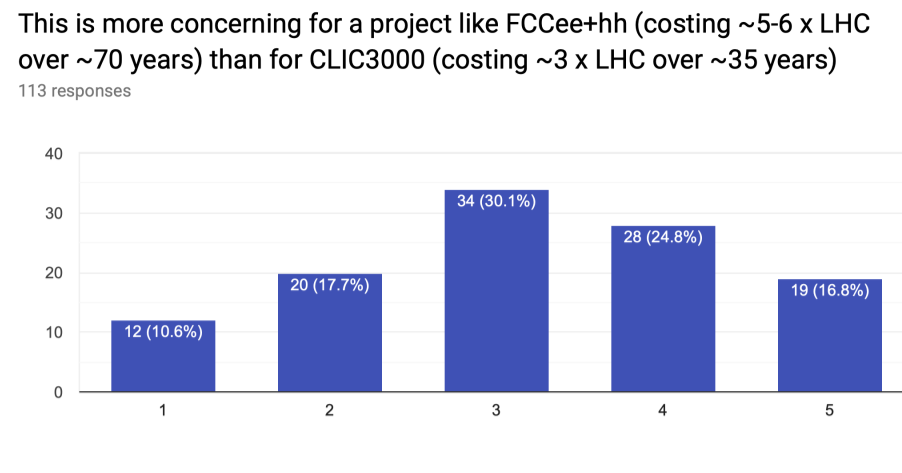}
    \end{center}
    \caption{Results of question 14 for the European Strategy Update section of the ECR survey, where (1) indicates strongly disagree, (2) disagree, (3) neutral, (4) agree and (5) strongly agree.}
    \label{fig:survey_ecfa_q14}
\end{figure}

\begin{figure}[h!]
    \begin{center}
    \includegraphics[width=0.9\textwidth]{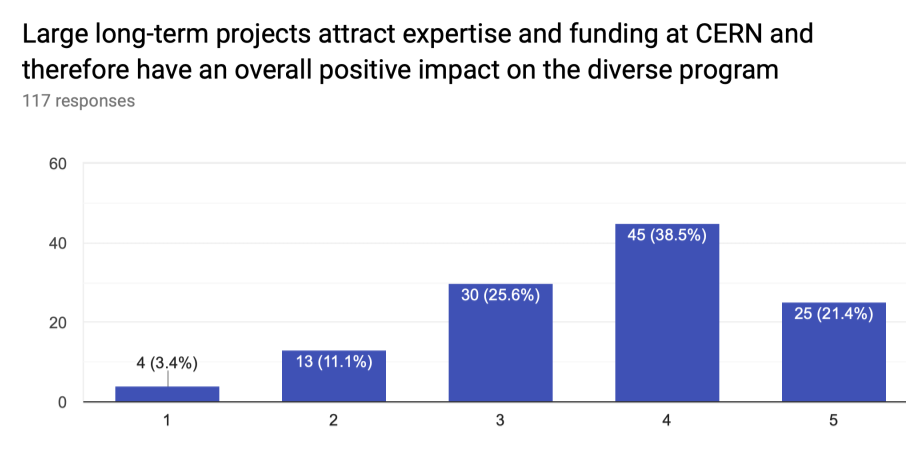}
    \end{center}
    \caption{Results of question 15 for the European Strategy Update section of the ECR survey, where (1) indicates strongly disagree, (2) disagree, (3) neutral, (4) agree and (5) strongly agree.}
    \label{fig:survey_ecfa_q15}
\end{figure}

\clearpage

\subsection{Human and Social Factors}
\begin{figure}[h!]
    \begin{center}
    \includegraphics[width=0.9\textwidth]{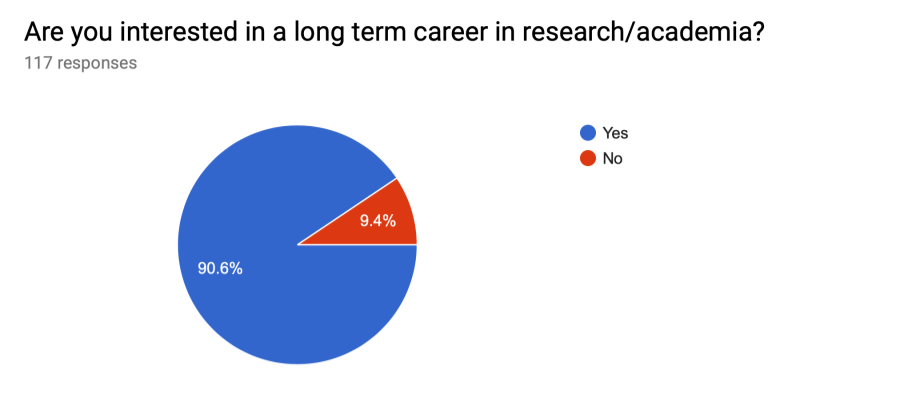}
    \end{center}
    \caption{Results of question 1 for the Human and Social Factors section of the ECR survey.}
    \label{fig:survey_humansocial_q1}
\end{figure}

\begin{figure}[h!]
    \begin{center}
    \includegraphics[width=0.9\textwidth]{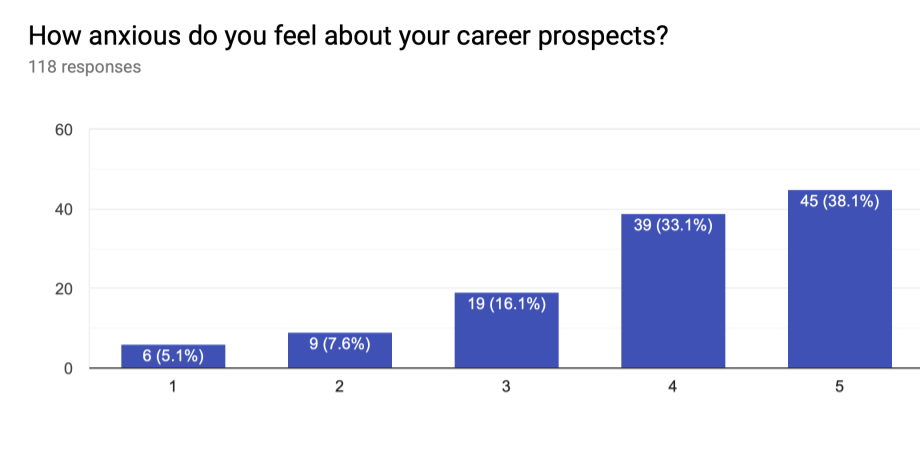}
    \end{center}
    \caption{Results of question 2 for the Human and Social Factors section of the ECR survey, where (1) indicates strongly disagree, (2) disagree, (3) neutral, (4) agree and (5) strongly agree.}
    \label{fig:survey_humansocial_q2}
\end{figure}

\begin{figure}[h!]
    \begin{center}
    \includegraphics[width=0.9\textwidth]{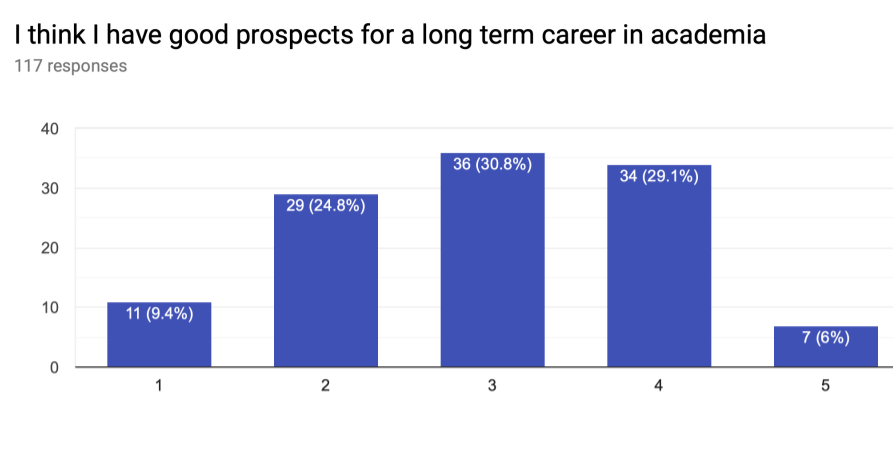}
    \end{center}
    \caption{Results of question 3 for the Human and Social Factors section of the ECR survey, where (1) indicates strongly disagree, (2) disagree, (3) neutral, (4) agree and (5) strongly agree.}
    \label{fig:survey_humansocial_q3}
\end{figure}

\begin{figure}[h!]
    \begin{center}
    \includegraphics[width=0.9\textwidth]{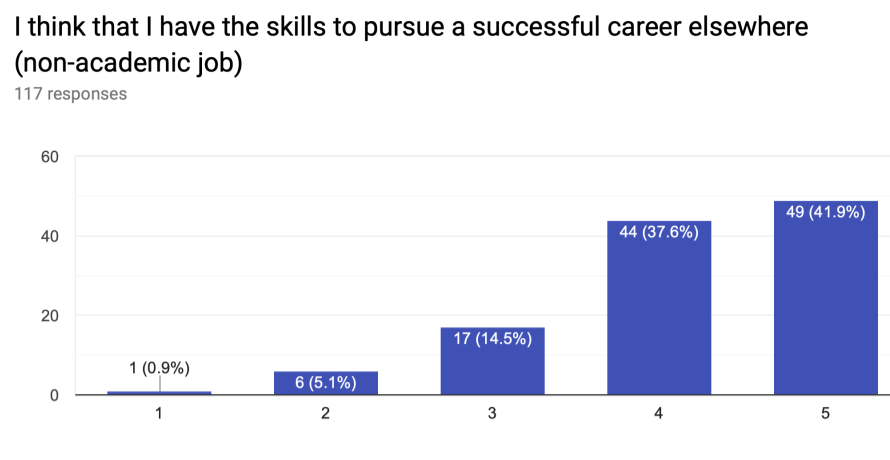}
    \end{center}
    \caption{Results of question 4 for the Human and Social Factors section of the ECR survey, where (1) indicates strongly disagree, (2) disagree, (3) neutral, (4) agree and (5) strongly agree.}
    \label{fig:survey_humansocial_q4}
\end{figure}

\begin{figure}[h!]
    \begin{center}
    \includegraphics[width=0.9\textwidth]{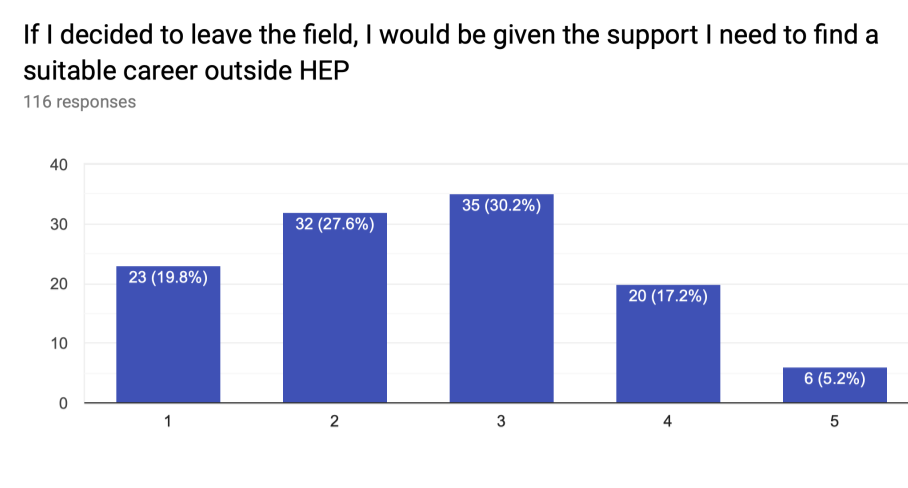}
    \end{center}
    \caption{Results of question 5 for the Human and Social Factors section of the ECR survey, where (1) indicates strongly disagree, (2) disagree, (3) neutral, (4) agree and (5) strongly agree.}
    \label{fig:survey_humansocial_q5}
\end{figure}

\begin{figure}[h!]
    \begin{center}
    \includegraphics[width=0.9\textwidth]{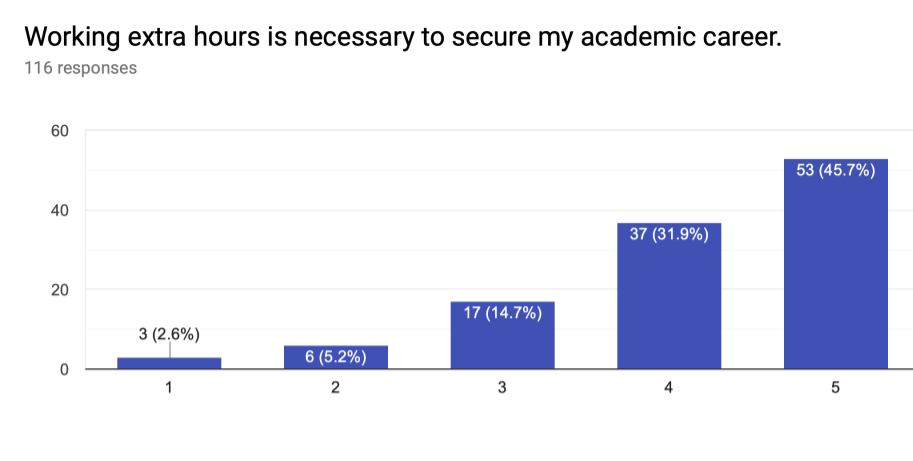}
    \end{center}
    \caption{Results of question 6 for the Human and Social Factors section of the ECR survey, where (1) indicates strongly disagree, (2) disagree, (3) neutral, (4) agree and (5) strongly agree.}
    \label{fig:survey_humansocial_q6}
\end{figure}

\begin{figure}[h!]
    \begin{center}
    \includegraphics[width=0.9\textwidth]{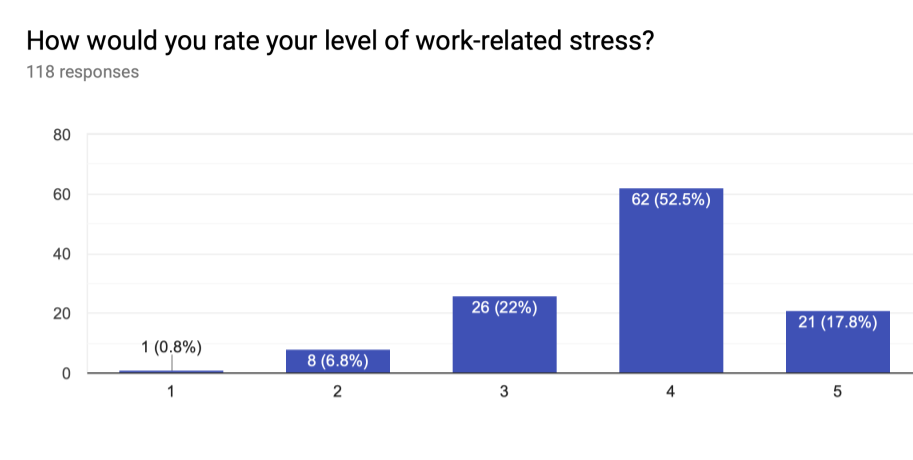}
    \end{center}
    \caption{Results of question 7 for the Human and Social Factors section of the ECR survey, where (1) indicates strongly disagree, (2) disagree, (3) neutral, (4) agree and (5) strongly agree.}
    \label{fig:survey_humansocial_q7}
\end{figure}

\begin{figure}[h!]
    \begin{center}
    \includegraphics[width=0.9\textwidth]{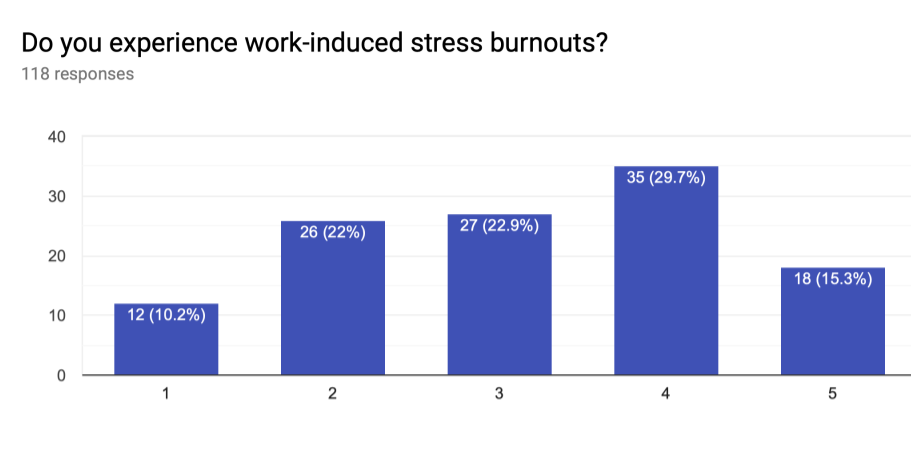}
    \end{center}
    \caption{Results of question 8 for the Human and Social Factors section of the ECR survey, where (1) indicates strongly disagree, (2) disagree, (3) neutral, (4) agree and (5) strongly agree.}
    \label{fig:survey_humansocial_q8}
\end{figure}
\begin{figure}[h!]
    \begin{center}
    \includegraphics[width=0.9\textwidth]{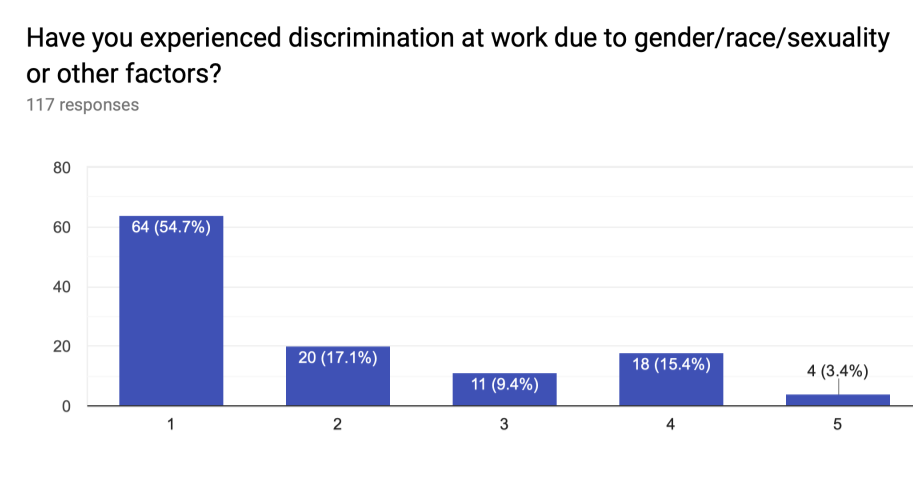}
    \end{center}
    \caption{Results of question 9 for the Human and Social Factors section of the ECR survey, where (1) indicates strongly disagree, (2) disagree, (3) neutral, (4) agree and (5) strongly agree.}
    \label{fig:survey_humansocial_q9}
\end{figure}

\begin{figure}[h!]
    \begin{center}
    \includegraphics[width=0.9\textwidth]{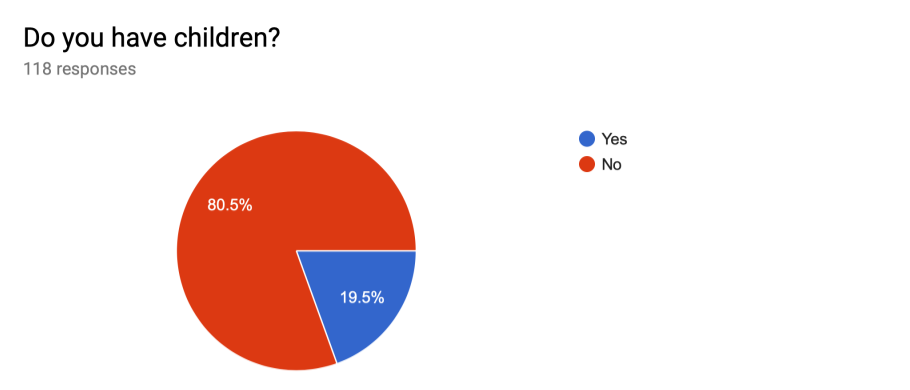}
    \end{center}
    \caption{Results of question 10 for the Human and Social Factors section of the ECR survey.}
    \label{fig:survey_humansocial_q10}
\end{figure}

\begin{figure}[h!]
    \begin{center}
    \includegraphics[width=0.9\textwidth]{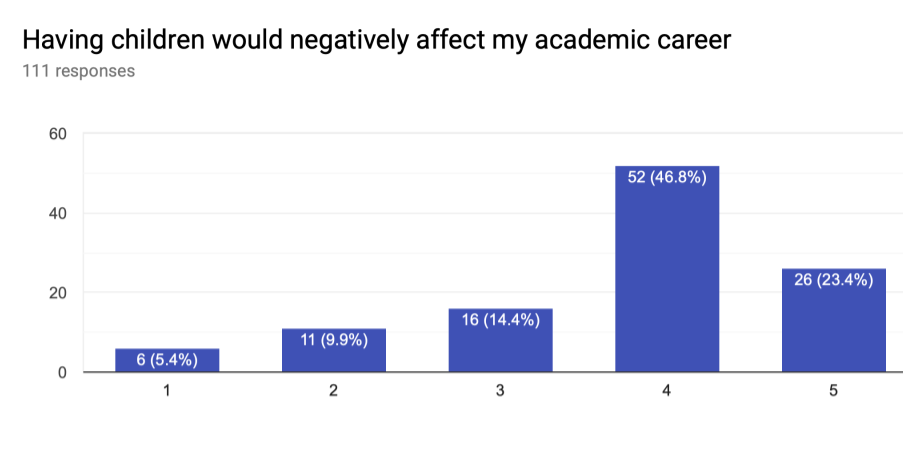}
    \end{center}
    \caption{Results of question 11 for the Human and Social Factors section of the ECR survey, where (1) indicates strongly disagree, (2) disagree, (3) neutral, (4) agree and (5) strongly agree.}
    \label{fig:survey_humansocial_q11}
\end{figure}

\begin{figure}[h!]
    \begin{center}
    \includegraphics[width=0.9\textwidth]{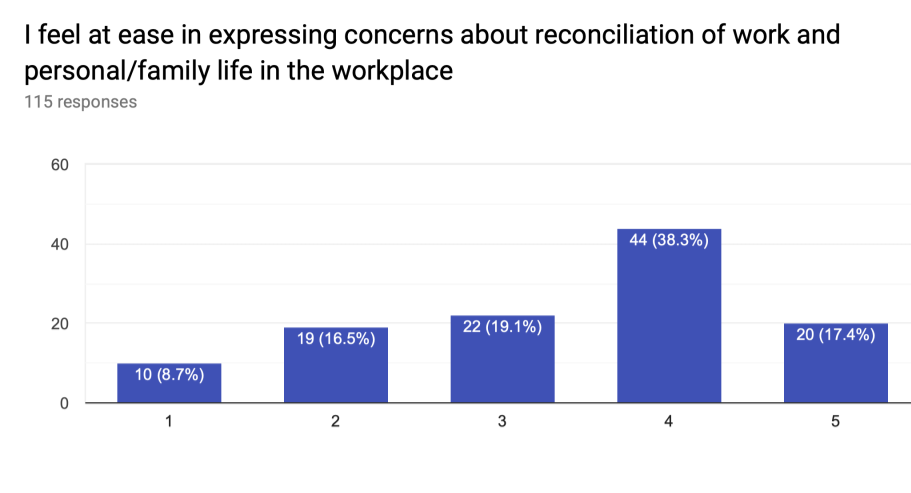}
    \end{center}
    \caption{Results of question 12 for the Human and Social Factors section of the ECR survey, where (1) indicates strongly disagree, (2) disagree, (3) neutral, (4) agree and (5) strongly agree.}
    \label{fig:survey_humansocial_q12}
\end{figure}

\begin{figure}[h!]
    \begin{center}
    \includegraphics[width=0.9\textwidth]{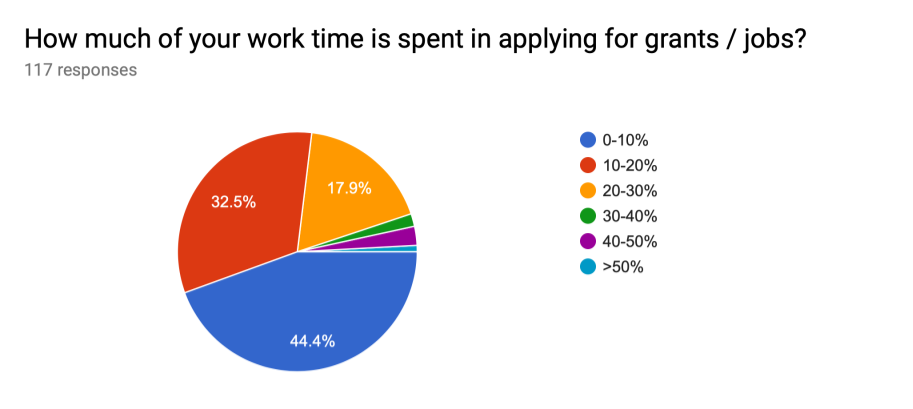}
    \end{center}
    \caption{Results of question 13 for the Human and Social Factors section of the ECR survey.}
    \label{fig:survey_humansocial_q13}
\end{figure}

\begin{figure}[h!]
    \begin{center}
    \includegraphics[width=0.9\textwidth]{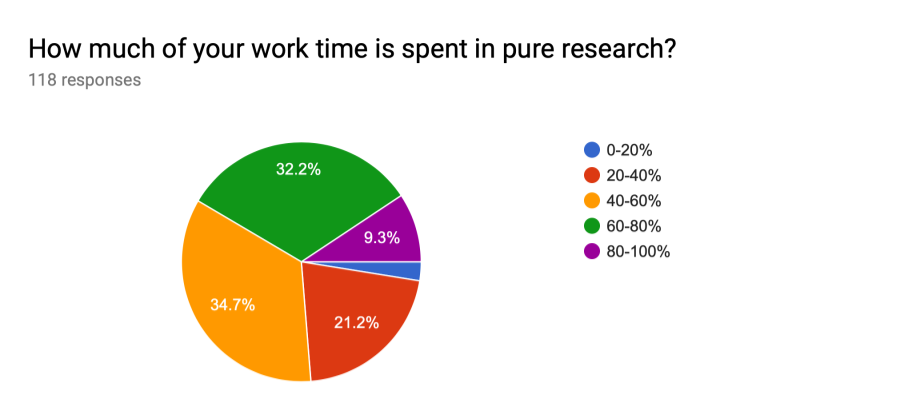}
    \end{center}
    \caption{Results of question 14 for the Human and Social Factors section of the ECR survey.}
    \label{fig:survey_humansocial_q14}
\end{figure}

\begin{figure}[h!]
    \begin{center}
    \includegraphics[width=0.9\textwidth]{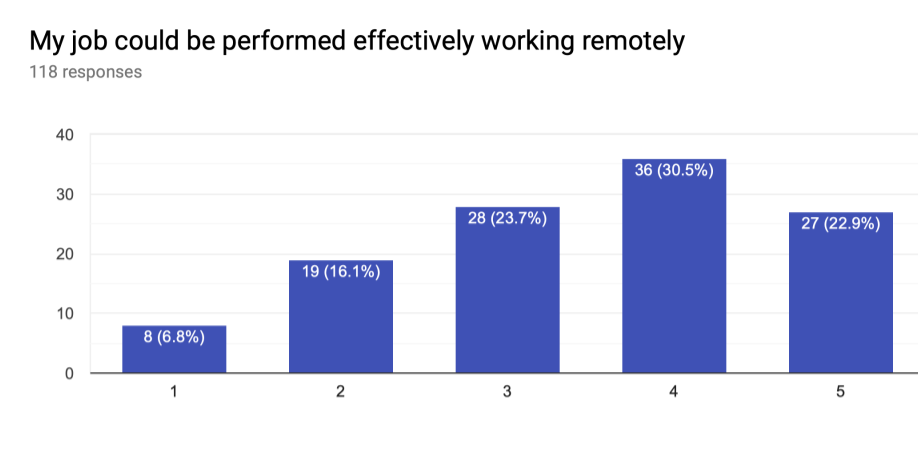}
    \end{center}
    \caption{Results of question 15 for the Human and Social Factors section of the ECR survey, where (1) indicates strongly disagree, (2) disagree, (3) neutral, (4) agree and (5) strongly agree.}
    \label{fig:survey_humansocial_q15}
\end{figure}

\begin{figure}[h!]
    \begin{center}
    \includegraphics[width=0.9\textwidth]{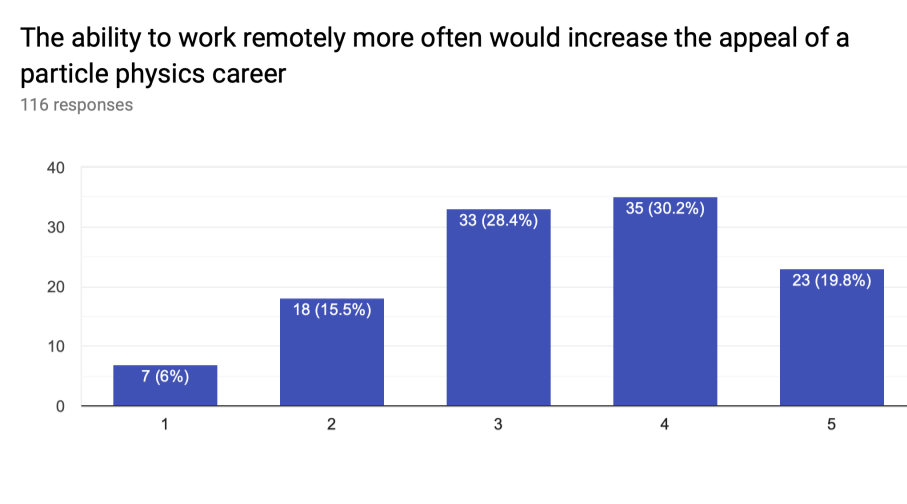}
    \end{center}
    \caption{Results of question 16 for the Human and Social Factors section of the ECR survey, where (1) indicates strongly disagree, (2) disagree, (3) neutral, (4) agree and (5) strongly agree.}
    \label{fig:survey_humansocial_q16}
\end{figure}

\begin{figure}[h!]
    \begin{center}
    \includegraphics[width=0.9\textwidth]{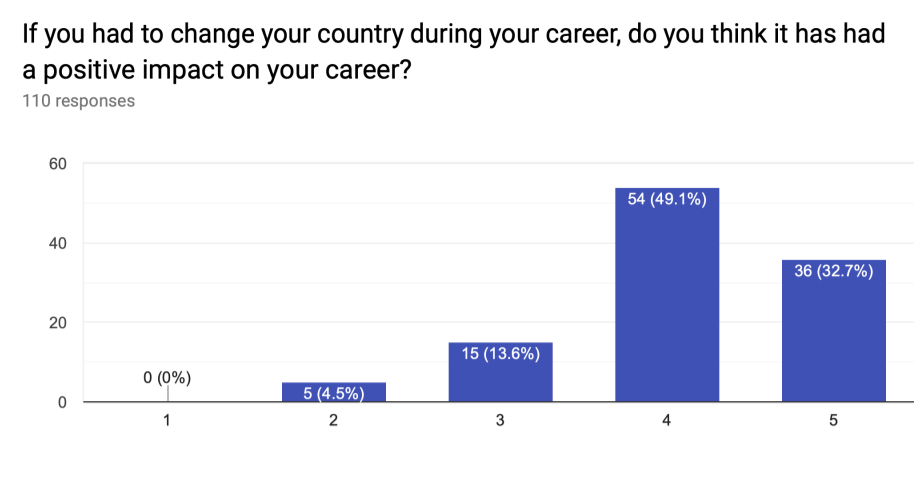}
    \end{center}
    \caption{Results of question 17 for the Human and Social Factors section of the ECR survey, where (1) indicates strongly disagree, (2) disagree, (3) neutral, (4) agree and (5) strongly agree.}
    \label{fig:survey_humansocial_q17}
\end{figure}

\begin{figure}[h!]
    \begin{center}
    \includegraphics[width=0.9\textwidth]{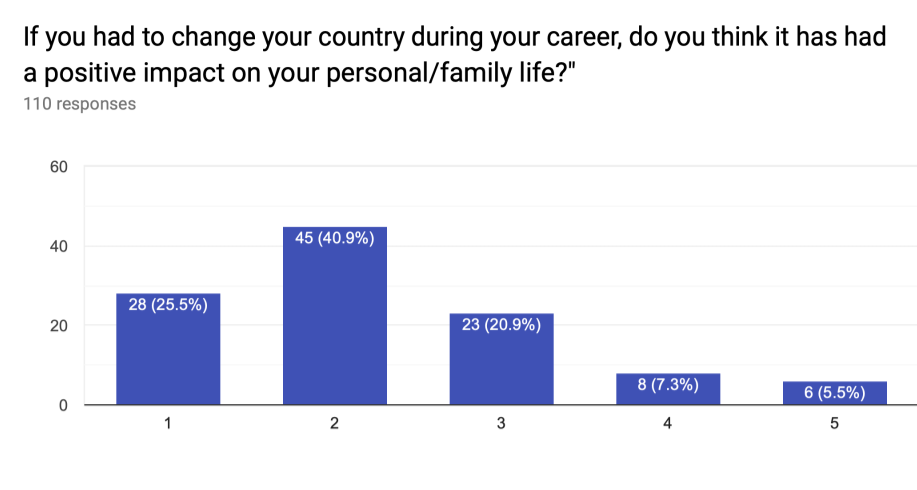}
    \end{center}
    \caption{Results of question 18 for the Human and Social Factors section of the ECR survey, where (1) indicates strongly disagree, (2) disagree, (3) neutral, (4) agree and (5) strongly agree.}
    \label{fig:survey_humansocial_q18}
\end{figure}

\clearpage

\subsection{Environment/Sustainability}

\begin{figure}[h!]
    \begin{center}
    \includegraphics[width=0.9\textwidth]{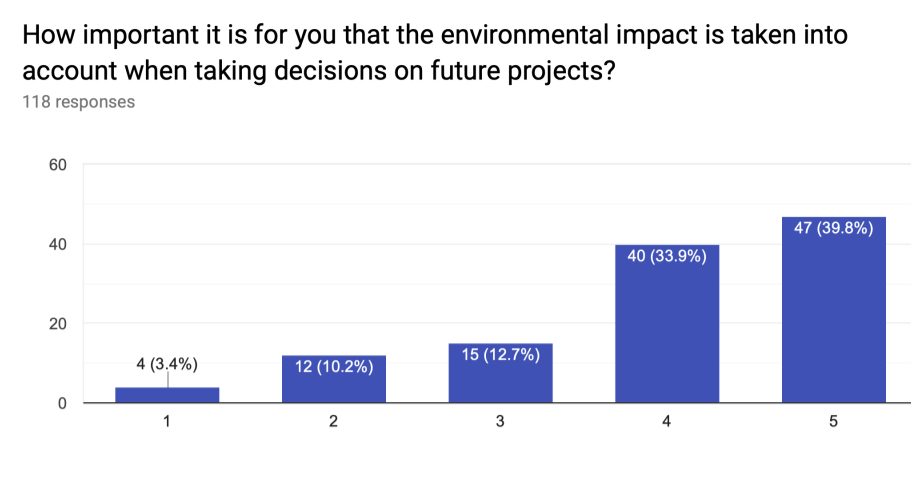}
    \end{center}
    \caption{Results of question 1 of the environment and sustainability section of the ECR survey, where (1) indicates strongly disagree, (2) disagree, (3) neutral, (4) agree and (5) strongly agree.}
    \label{fig:survey_env_q1}
\end{figure}

\begin{figure}[h!]
    \begin{center}
    \includegraphics[width=0.9\textwidth]{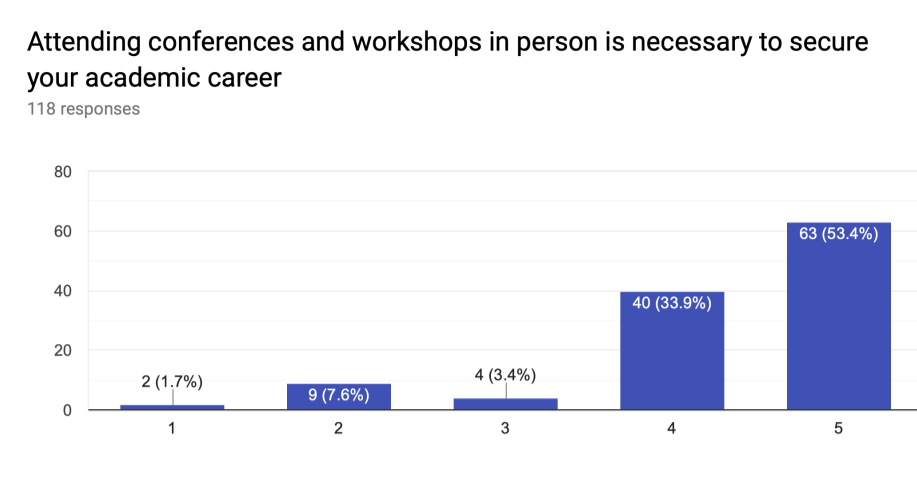}
    \end{center}
    \caption{Results of question 2 of the environment and sustainability section of the ECR survey, where (1) indicates strongly disagree, (2) disagree, (3) neutral, (4) agree and (5) strongly agree.}
    \label{fig:survey_env_q2}
\end{figure}

\begin{figure}[h!]
    \begin{center}
    \includegraphics[width=0.9\textwidth]{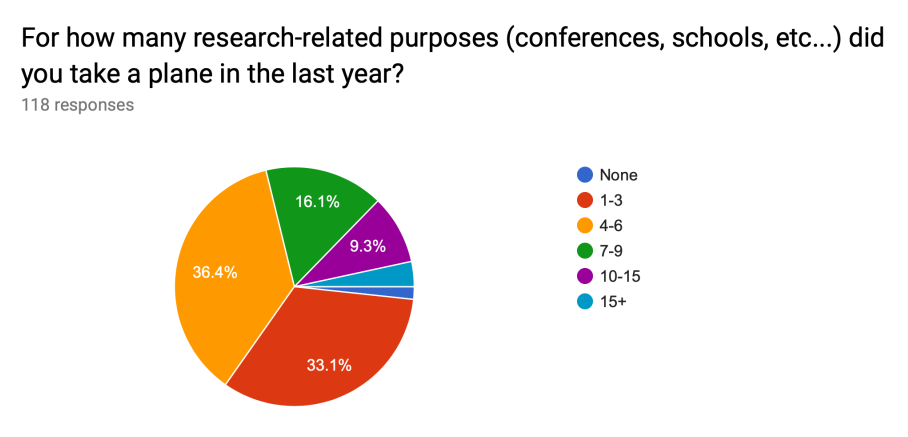}
    \end{center}
    \caption{Results of question 3 of the environment and sustainability section of the ECR survey.}
    \label{fig:survey_env_q3}
\end{figure}

\begin{figure}[h!]
    \begin{center}
    \includegraphics[width=0.9\textwidth]{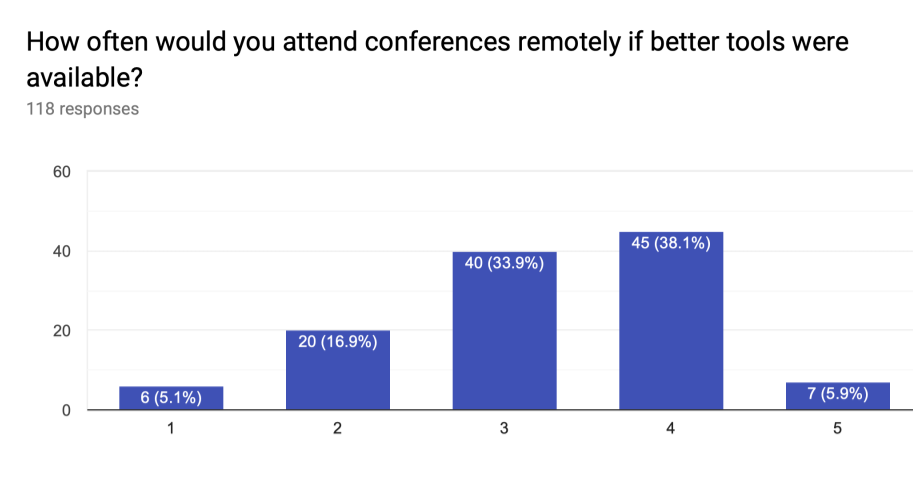}
    \end{center}
    \caption{Results of question 4 of the environment and sustainability section of the ECR survey, where (1) indicates strongly disagree, (2) disagree, (3) neutral, (4) agree and (5) strongly agree.}
    \label{fig:survey_env_q4}
\end{figure}

\begin{figure}[h!]
    \begin{center}
    \includegraphics[width=0.9\textwidth]{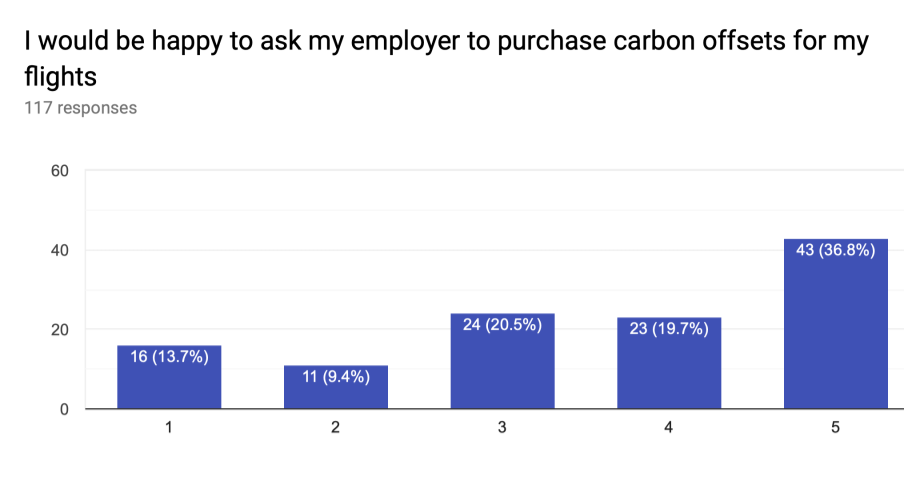}
    \end{center}
    \caption{Results of question 5 of the environment and sustainability section of the ECR survey, where (1) indicates strongly disagree, (2) disagree, (3) neutral, (4) agree and (5) strongly agree.}
    \label{fig:survey_env_q5}
\end{figure}

\begin{figure}[h!]
    \begin{center}
    \includegraphics[width=0.9\textwidth]{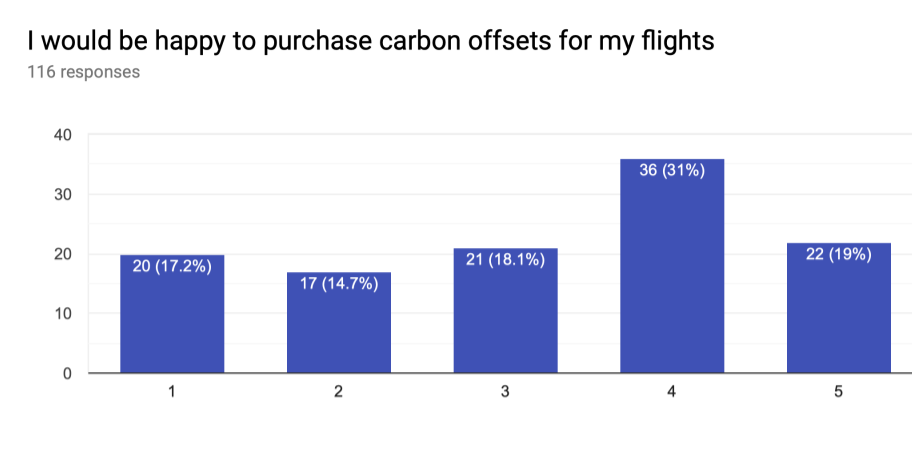}
    \end{center}
    \caption{Results of question 6 of the environment and sustainability section of the ECR survey, where (1) indicates strongly disagree, (2) disagree, (3) neutral, (4) agree and (5) strongly agree.}
    \label{fig:survey_env_q6}
\end{figure}

\begin{figure}[h!]
    \begin{center}
    \includegraphics[width=0.9\textwidth]{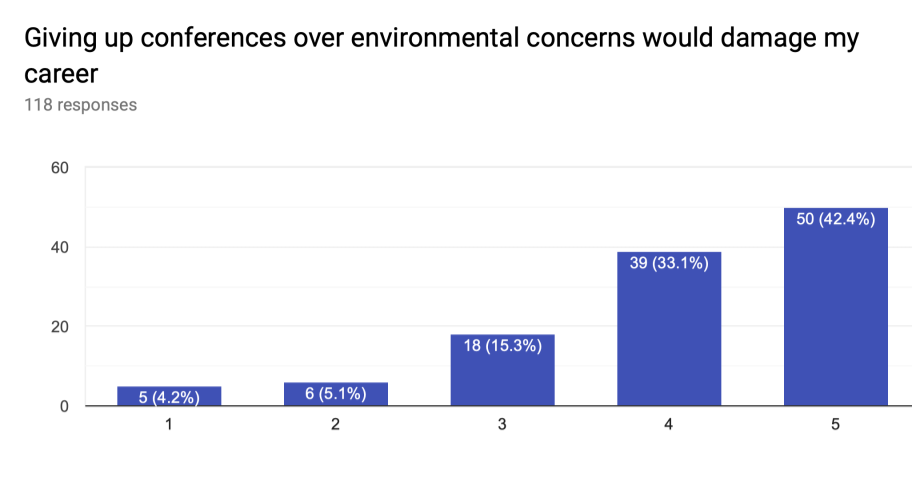}
    \end{center}
    \caption{Results of question 7 of the environment and sustainability section of the ECR survey, where (1) indicates strongly disagree, (2) disagree, (3) neutral, (4) agree and (5) strongly agree.}
    \label{fig:survey_env_q7}
\end{figure}

\begin{figure}[h!]
    \begin{center}
    \includegraphics[width=0.9\textwidth]{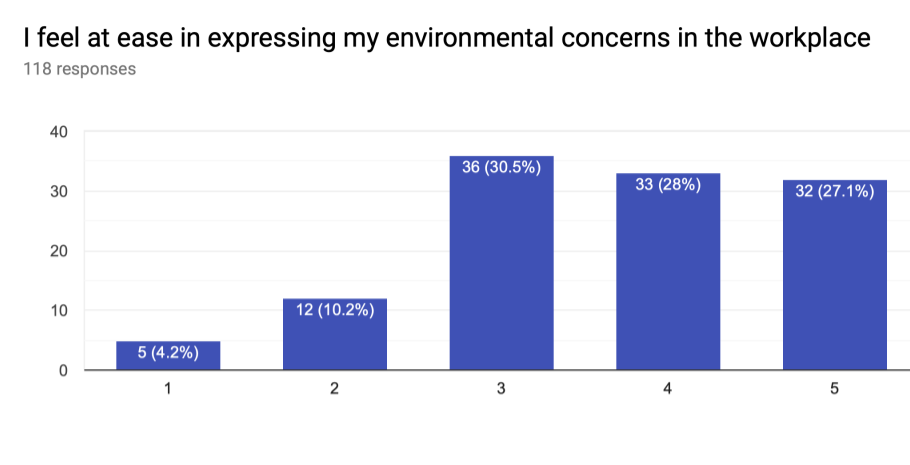}
    \end{center}
    \caption{Results of question 8 of the environment and sustainability section of the ECR survey, where (1) indicates strongly disagree, (2) disagree, (3) neutral, (4) agree and (5) strongly agree.}
    \label{fig:survey_env_q8}
\end{figure}

\begin{figure}[h!]
    \begin{center}
    \includegraphics[width=0.9\textwidth]{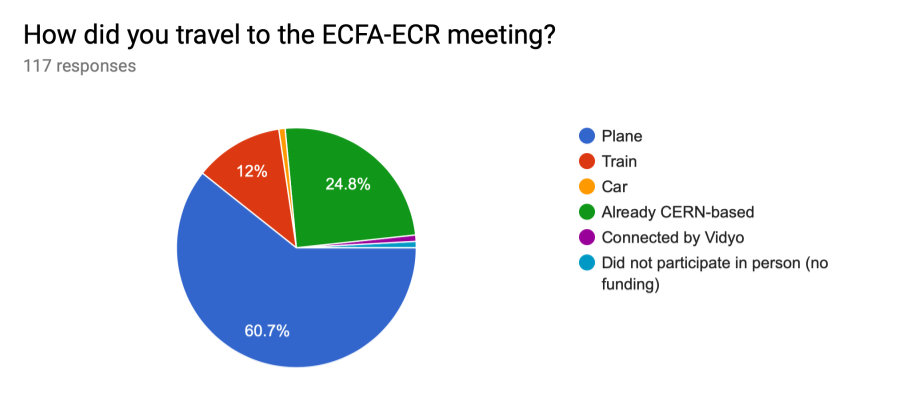}
    \end{center}
    \caption{Results of question 9 of the environment and sustainability section of the ECR survey.}
    \label{fig:survey_env_q9}
\end{figure}

\end{document}